\documentclass[a4paper,11pt,oneside]{article}

\usepackage[english]{babel}
\usepackage[T1]{fontenc}
\usepackage[utf8]{inputenc}

\usepackage[pdftex,unicode]{hyperref}
\hypersetup{pdftitle=Modeling Trade Durations under Temporal Granularity Effects in Forex Markets}
\hypersetup{pdfauthor=Vladimír Holý}

\usepackage{tikz}

\usepackage{xcolor}
\definecolor{mycol}{rgb}{0.07,0.14,0.35}
\hypersetup{colorlinks=true, linkcolor=mycol, anchorcolor=mycol, citecolor=mycol, filecolor=mycol, urlcolor=mycol}

\usepackage[margin=60pt]{geometry}
\usepackage[authoryear]{natbib}

\usepackage{amsmath}
\usepackage{amssymb}
\usepackage{graphicx}
\usepackage[group-minimum-digits=3]{siunitx}
\usepackage{booktabs}
\usepackage{multirow}
\usepackage{tablefootnote}
\usepackage{float}
\usepackage{caption}

\usepackage[normalem]{ulem}

\usepackage{pifont}

\usepackage[super]{nth}

\begin{document}

\begin{center}
{\Large \bfseries Modeling Trade Durations under Temporal Granularity Effects \\ in Forex Markets}
\end{center}

\begin{center}
{\bfseries Vladimír Holý} \\
Prague University of Economics and Business \\
Winston Churchill Square 4, 130 67 Prague 3, Czechia \\
\href{mailto:vladimir.holy@vse.cz}{vladimir.holy@vse.cz} \\
\end{center}

\noindent
\textbf{Abstract:}
Trade durations in high-frequency foreign exchange data exhibit increased occurrence near integer values. To address this empirical phenomenon, we propose the granularity-adjusted autoregressive conditional duration (GA-ACD) model. It is based on a novel two-component mixture distribution consisting of a standard generalized gamma component for regular durations and a second component that locally redistributes probability mass around integer values to capture heaping. Conditional dynamics are modeled within a score-driven framework, allowing the scale parameter to vary over time in response to past durations, and enabling maximum likelihood estimation of all model parameters. A simulation study shows that ignoring heaping leads to biased parameter estimates and distorted inference regarding both the distribution and the dynamics of durations. An empirical analysis demonstrates that integer-duration clustering is pervasive across major currency pairs and that the GA-ACD model outperforms the standard generalized gamma ACD model.
\\

\\

\\

\section{Introduction}
\label{sec:intro}

The autoregressive conditional duration (ACD) model, introduced by \cite{Engle1998}, provides a framework for modeling the time intervals between events in high-frequency financial data. The ACD model captures clustering and time-varying dynamics in event arrivals, making it well suited for market microstructure analysis. The original specification assumed the exponential distribution for the error term, but numerous alternative distributions have since been proposed to provide greater flexibility in modeling duration dynamics (see \citealp{Pacurar2008} and \citealp{Saranjeet2019}).

\begin{figure}
\centering
\includegraphics[scale=0.6]{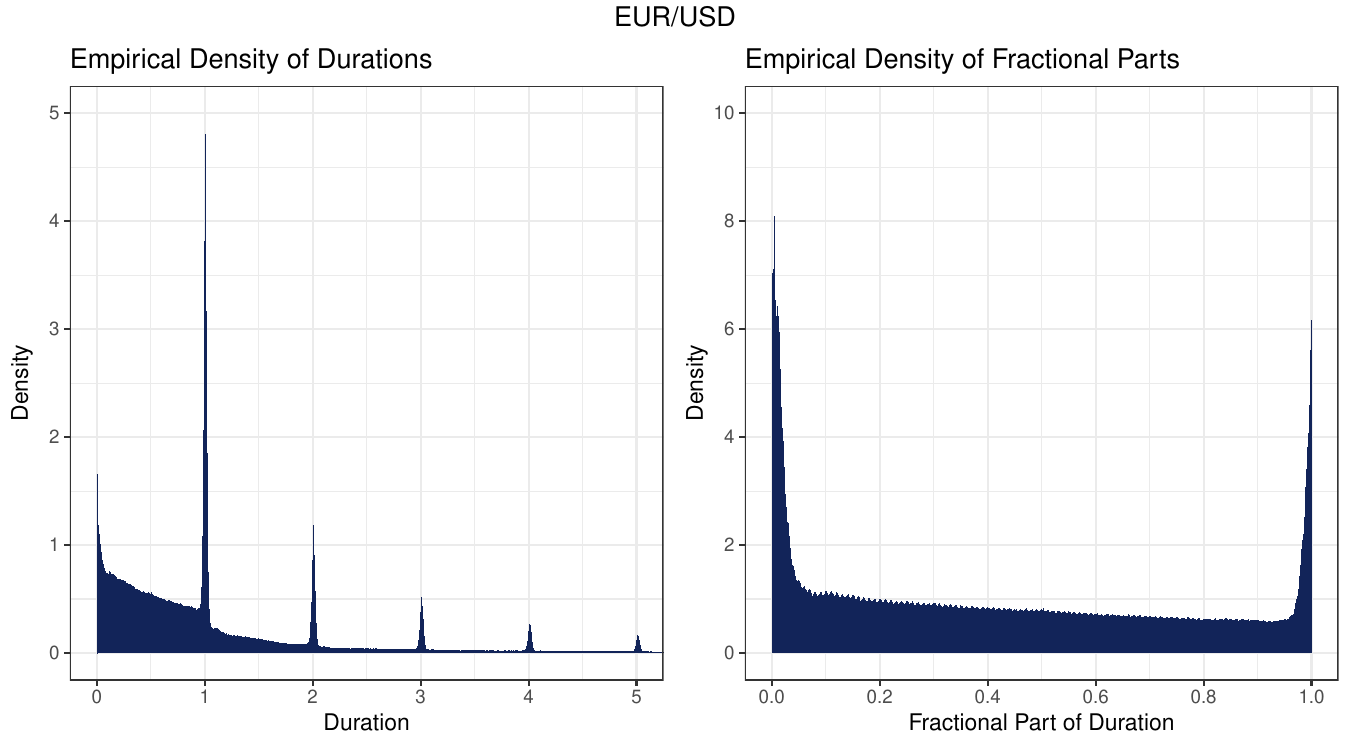}
\caption{Empirical densities of trade durations and their fractional part for the EUR/USD pair.}
\label{fig:duration_eur}
\end{figure}

This paper focuses on a prominent empirical feature of foreign exchange (FX) data: trade durations exhibit excessive clustering at and around integer values, a phenomenon referred to as heaping. This behavior is illustrated in Figure \ref{fig:duration_eur}. The left panel shows pronounced clustering of durations around $1, 2, 3, \ldots$ seconds, resulting in elevated density near these integer values. Importantly, the clustering is not confined to exact integers but also extends to durations in their immediate vicinity. This pattern is evident in the right panel, which displays the density of the fractional parts of duration values and reveals a clear concentration near 0 and 1 thresholds. Notably, the heaping phenomenon is observed in trade durations, but not in the transaction timestamps themselves. Furthermore, heaping appears to be specific to FX markets and is not observed, for example, in stock markets.

A plausible explanation for the observed heaping is not trading behavior itself, but rather the way FX transaction data are reported and consolidated. FX data vendors typically combine transaction records from multiple trading venues, which may differ in timestamp precision, clock synchronization, and other reporting conventions. In particular, some venues record transactions with only second-level precision, whereas others provide timestamps at a finer resolution. When such heterogeneous data streams are combined, additional discrepancies may arise from imperfect synchronization of system clocks and from the aggregation procedures used by data vendors. These factors can generate excessive clustering of duration values around integer seconds, even when the underlying transaction times do not exhibit such clustering.
 
Because standard continuous-duration models do not account for the heaping, they may yield biased parameter estimates and a distorted representation of both the duration distribution and its dynamics. To address this, we propose a novel distribution for FX trade durations that explicitly accounts for granularity-induced heaping. The model is constructed as a two-component mixture: standard durations are modeled using the generalized gamma distribution, while durations concentrated near integer values are captured by redistributing the corresponding generalized gamma probability mass around integer points via the truncated normal distribution. The conditional dynamics are specified within a score-driven framework of \cite{Creal2013} and \cite{Harvey2013}, allowing time-varying parameters to be driven by the score, i.e. the gradient of the log-likelihood. The model can be estimated by the maximum likelihood method. We refer to the resulting specification as the granularity-adjusted autoregressive conditional duration (GA-ACD) model.

Several extensions of the ACD framework employ mixture distributions to better capture empirical features of duration data. For example, \cite{DeLuca2004} proposed a mixture of two exponential distributions to provide greater flexibility in modeling the shape and tail behavior of durations. This approach was further extended by \cite{Yatigammana2019}, who developed more general mixture-based specifications within the ACD setting. \cite{Blasques2024a} and \cite{Shi2021} addressed the excessive occurrence of zero and close-to-zero durations---associated with split transactions in stock markets---by using zero-inflated distributions within the ACD setting. However, none of these studies addresses the specific form of heaping observed in FX markets.

A somewhat related phenomenon---price clustering---occurs in stock markets, where transaction prices disproportionately concentrate at round tick values, such as multiples of 5 or 10 cents, and has spawned a sizeable literature. For instance, \cite{Holy2022b} developed a GARCH-like model based on the double Poisson distribution with inflated probability mass at 5 and 10 cents to capture such clustering. Crucially, price clustering is fundamentally linked to the discrete nature of stock prices and is driven by the behavior of traders, whereas clustering in FX trade durations arises in an inherently continuous setting and is attributable to data processing mechanisms, such as aggregation, rounding, and reporting conventions, rather than to intrinsic discreteness.

The rest of the paper is structured as follows. In Section \ref{sec:model}, we introduce the granularity-adjusted ACD model. In Section \ref{sec:sim}, we present the simulation study comparing the proposed model with the model that ignores heaping. In Section \ref{sec:emp}, we provide the empirical analysis of seven major FX currency pairs. We conclude the paper in Section \ref{sec:con}. Additional details on the model are provided in Appendix \ref{app:model}, additional simulation results are reported in Appendix \ref{app:sim}, and additional empirical evidence is presented in Appendix \ref{app:emp}.

\section{Granularity-Adjusted ACD Model}
\label{sec:model}

Our main object of interest is the trade duration, i.e.\ the interval between consecutive transaction arrival times. For $i = 1, \ldots, n$, we define the random trade duration as $X_i = T_i - T_{i-1}$ and its observed value as $x_i = t_i - t_{i-1}$. First, we propose a novel mixture distribution for $X_i$, and then we specify dynamics for its time-varying scale parameter $\lambda_i$. Finally, we discuss estimation via the maximum likelihood method.

\subsection{Mixture Distribution}
\label{sec:modelDistr}

We model $X_i$ as a mixture of two components:
\begin{equation}
X_i = \begin{cases}
Y_i & \text{with probability } 1 - \rho, \\
Z_i & \text{with probability } \rho, \\
\end{cases}
\end{equation}
where $Y_i$ represents standard durations and $Z_i$ captures durations concentrated around integer values. The parameter $\rho \in [0,1]$ denotes the mixture weight. The density function of the mixture distribution is given by
\begin{equation}
\label{eq:densityMix}
f_{X_i}(x_i) = (1 - \rho) f_{Y_i}(x_i) + \rho f_{Z_i}(x_i).
\end{equation}

We assume that $Y_i$ follows the generalized gamma distribution with time-varying scale parameter $\lambda_i$ and two static shape parameters $\gamma$ and $\kappa$. Its density function is given by
\begin{equation}
\label{eq:densityGengamma}
f_{Y_i} (x_i) = \frac{1}{\Gamma(\gamma)} \frac{\kappa}{\lambda_i} \left( \frac{x_i}{\lambda_i} \right)^{\gamma \kappa - 1} \exp \left(- \left( \frac{x_i}{\lambda_i} \right)^{\kappa} \right).
\end{equation}

This parametrization reduces to the gamma distribution when $\kappa=1$, to the Weibull distribution when $\gamma=1$, and to the exponential distribution when $\gamma=\kappa=1$. Our specific choice of the generalized gamma distribution is not crucial for the proposed framework. Alternative distributions---such as the Birnbaum--Saunders distribution \citep{Bhatti2010}, the Burr distribution \citep{Grammig2000}, the Fréchet distribution \citep{Zheng2016}, the generalized F distribution \citep{Hautsch2001}, or the log-normal distribution \citep{Xu2013}---could be employed instead.

\begin{figure}
\centering
\includegraphics[scale=0.6]{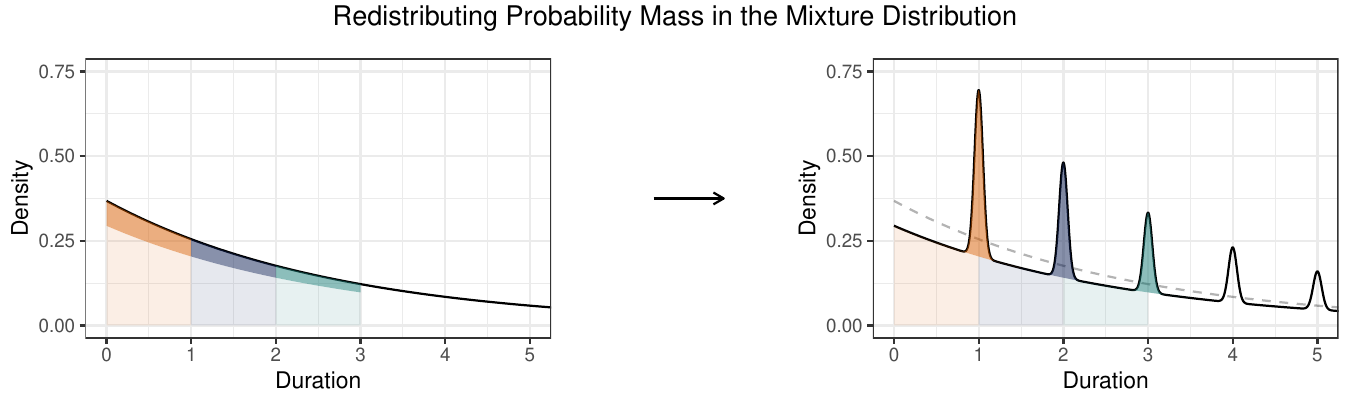}
\caption{Illustration of the redistribution of probability mass toward peaks at integer values. Parameters of the mixture distribution are $\lambda_i = 1$, $\gamma = 1$, $\kappa = 1$, $\rho = 0.20$, and $\sigma = 0.05$.}
\label{fig:redistr}
\end{figure}

We derive the distribution of $Z_i$ by locally concentrating the distribution for $Y_i$, with the same parameters $\lambda_i$, $\gamma$, and $\kappa$, around integer values. We take the probability mass on $[ \lfloor x_i \rceil - 1, \lfloor x_i \rceil )$ (given by the distribution for $Y_i$) and redistribute it to $[ \lfloor x_i \rceil - 0.5, \lfloor x_i \rceil + 0.5 )$ according to the (truncated) normal distribution with mean $\lfloor x_i \rceil$ and standard deviation $\sigma$. This process is illustrated in Figure \ref{fig:redistr}. The density function of $Z_i$ is then
\begin{equation}
\label{eq:densityConc}
f_{Z_i}(x_i) = \frac{ F_{Y_i}(\lfloor x_i \rceil) - F_{Y_i}(\lfloor x_i \rceil - 1) }{ F_{W_i}(\lfloor x_i \rceil + 0.5) - F_{W_i}(\lfloor x_i \rceil - 0.5) } f_{W_i}( x_i ), \qquad W_i \sim \mathrm{Normal}(\lfloor x_i \rceil, \sigma).
\end{equation}
This corresponds to rounding durations up to integer values and then introducing noise to timing using a normal distribution. Note that when $x_i < 0.5$, the rounded value is 0, which is assigned zero probability. Only values near positive integers are thus inflated, which is consistent with our empirical observations.

The resulting mixture distribution has five parameters: the scale parameter $\lambda_i > 0$, which determines the overall time scale (larger $\lambda_i$ implies longer expected durations); the two shape parameters $\gamma>0$ and $\kappa>0$, which jointly govern the behavior of the hazard function (allowing for decreasing, increasing, or non-monotonic hazard rates); the mixture weight parameter $\rho \in [0, 1]$, which controls the relative weight of the two components (larger $\rho$ assigns more probability mass to the component capturing durations concentrated around integer transaction times); and the spread parameter $\sigma > 0$, which regulates the dispersion of the concentration component (smaller $\sigma$ implies tighter clustering around integer values, whereas larger $\sigma$ leads to more dispersed durations).

\begin{figure}
\centering
\includegraphics[scale=0.6]{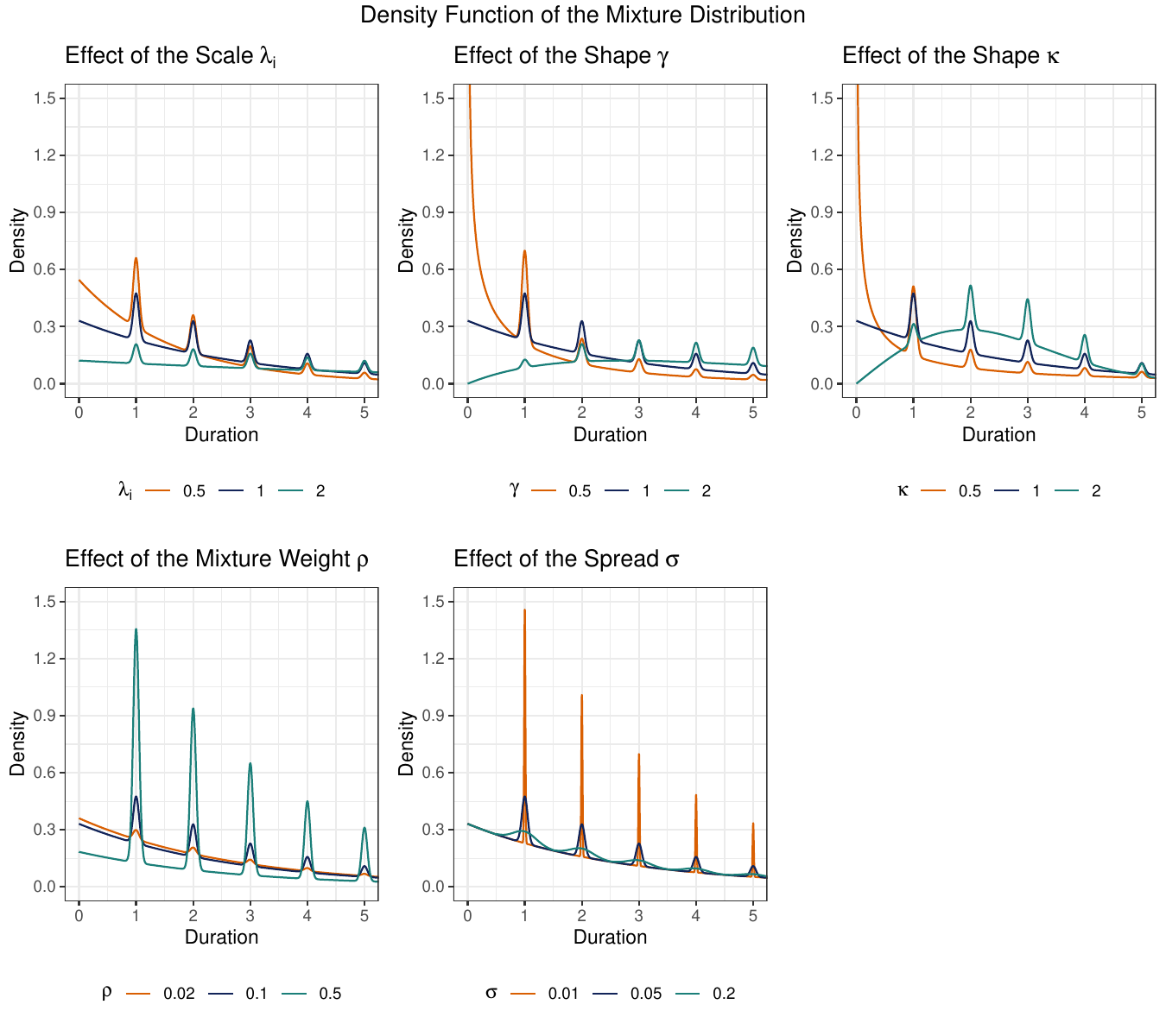}
\caption{The effects of parameter values on the density function of the mixture distribution. In each plot, a single parameter is varied while the others are held at their default values: $\lambda_i = 1$, $\gamma = 1$, $\kappa = 1$, $\rho = 0.10$, and $\sigma = 0.05$.}
\label{fig:distr}
\end{figure}

Figure \ref{fig:distr} illustrates the behavior of the density function with respect to its five parameters. Additional details on the mixture distribution are provided in Appendix \ref{app:model}.

\subsection{Scale Dynamics}
\label{sec:modelDyn}

Similarly to the ACD literature, we allow the scale parameter to be time-varying. As $\lambda_i$ is required to be positive, we specify its dynamics in terms of $\ln \lambda_i$, which does not require additional constraints on the parameters. We assume that the scale is influenced by two main components: deterministic seasonal pattern $S_i$ and stochastic time dependence $E_i$.

The seasonal component captures recurring intraday and intraweek effects typical in the forex market, such as heightened activity during overlapping trading sessions. We estimate this component nonparametrically using smoothing splines, fitted to the log-transformed durations as a function of time-of-week. This approach captures the systematic variation in trading activity across hours and days while avoiding the imposition of a parametric form. We do not adjust durations for this seasonal pattern beforehand; instead, we include the seasonal component directly in the dynamic model to avoid distorting clustering near integer values.

The time-dependent component accounts for the persistence in durations, allowing the scale of the distribution to adjust dynamically in response to recent market activity. We model this within the framework of score-driven (GAS) models of \cite{Creal2013} and \cite{Harvey2013}. Specifically, the dynamics are given by
\begin{equation}
\label{eq:dyn}
\ln \lambda_i = \omega + S_i + E_i, \qquad E_i = \varphi E_{i-1} + \alpha \nabla_{Y_i; \ln \lambda_i}(x_i),
\end{equation}
where $\omega$ is the constant parameter, $\varphi$ is the autoregressive parameter governing persistence, and $\alpha$ is the score parameter controlling the response to new information. The score is the gradient of the log-likelihood, i.e.
\begin{equation}
\label{eq:score}
\nabla_{X_i; \ln \lambda_i}(x_i) = \frac{\partial \ln f_{X_i} (x_i)}{\partial \ln \lambda_i}.
\end{equation}
For the proposed mixture distribution based on the generalized gamma distribution, it is derived in Appendix \ref{app:model}. The score provides the direction and magnitude of the steepest ascent in likelihood given new observations, allowing the model to update the conditional scale efficiently in response to recent market activity. Compared to standard ACD models, the score-driven approach adapts automatically to the local shape of the conditional distribution, improving responsiveness to extreme events and capturing short-term clustering more accurately.

\subsection{Maximum Likelihood Estimation}
\label{sec:modelMle}

We estimate the parameters of the proposed model $f = (\omega, \varphi, \alpha, \gamma, \kappa, \rho, \sigma)^{\intercal}$ by the maximum likelihood method. The likelihood is constructed from the conditional density of the mixture distribution, with the time-varying scale updated recursively. All parameters are jointly estimated via numerical optimization. Standard errors of the parameter estimates can be obtained from the inverse of the empirical Hessian of the log-likelihood evaluated at the maximum likelihood estimates. For the theoretical properties of maximum likelihood estimation in score-driven models, we refer to \cite{Blasques2014a}, \cite{Blasques2018}, and \cite{Blasques2022}.

\section{Simulation Study}
\label{sec:sim}

The Monte Carlo study serves two purposes. First, it demonstrates that the standard score-driven ACD model based on the generalized gamma distribution yields biased parameter estimates and distorted inference regarding both the distribution and the dynamics of durations when heaping is present but ignored. Second, it evaluates the finite-sample performance of the proposed GA-ACD model and its ability to recover the underlying data-generating process. This section presents the main findings, while additional results are provided in Appendix \ref{app:sim}.

Data are generated from the GA-ACD model with parameter values $\omega = 0$, $\alpha = 0.25$, $\varphi = 0.998$, $\gamma = 1.2$, $\kappa = 0.8$, $\rho = 0.2$, and $\sigma = 0.015$. These values are consistent with those obtained in the empirical analysis presented in Section \ref{sec:emp}. Unless stated otherwise, the sample size is set to $n = 10\,000$. In selected simulation scenarios, either $\rho$ or $n$ is varied to examine the sensitivity of the results to the degree of heaping and the sample size. For each parameter configuration, 2\,000 Monte Carlo replications are generated.

\subsection{Misspecification of the standard ACD Model}
\label{sec:simBias}

\begin{figure}
\centering
\includegraphics[scale=0.6]{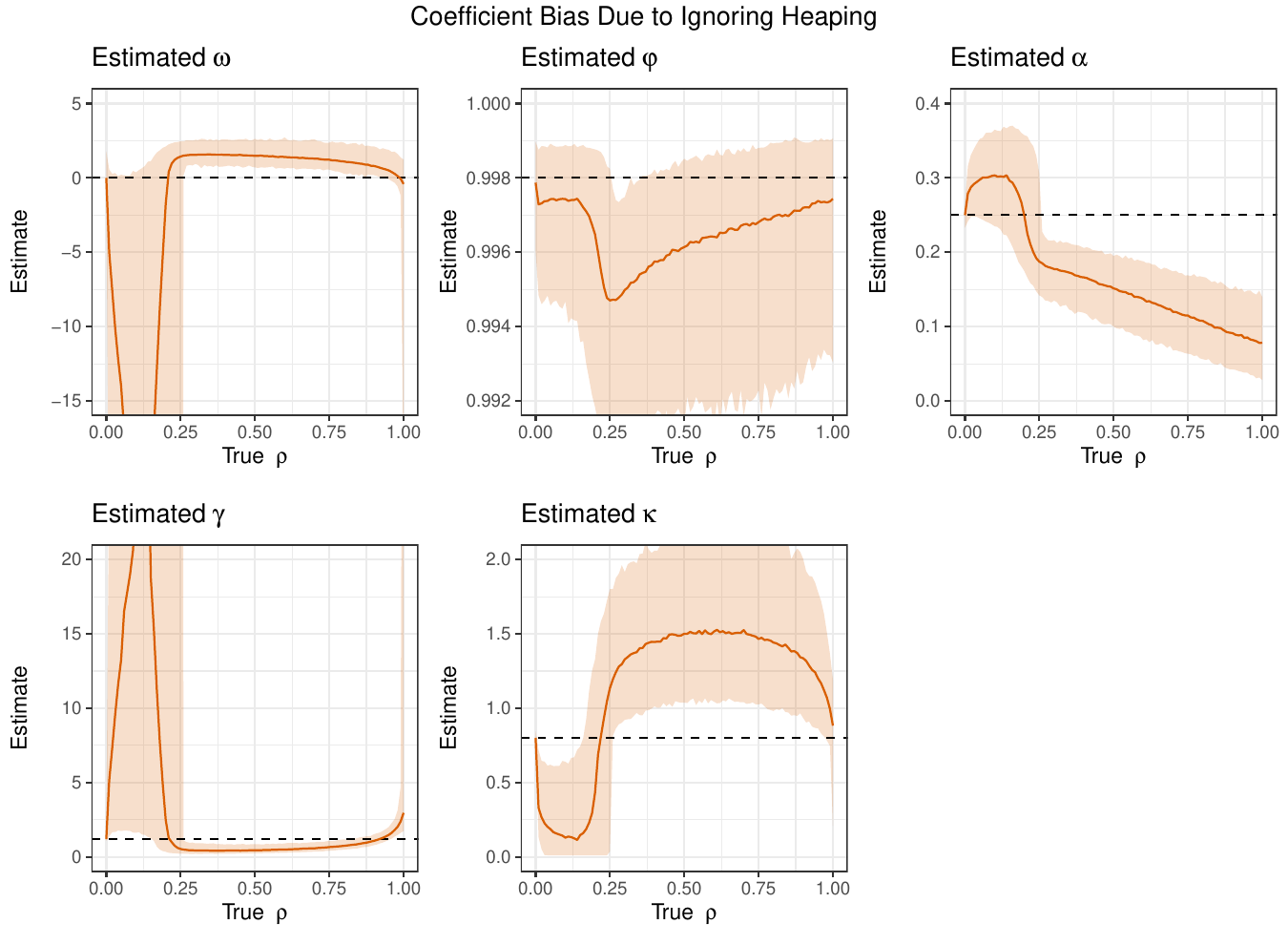}
\caption{Median parameter estimates together with the 95\% interval across simulation replications for the standard ACD model. Results are based on 2,000 Monte Carlo replications.}
\label{fig:coefBias}
\end{figure}

Figure \ref{fig:coefBias} shows that ignoring heaping leads to substantial and systematic distortions in the estimated parameters. These distortions arise even for relatively small values of $\rho$, when the degree of heaping is limited. Both the magnitude and the direction of the bias vary nonlinearly with $\rho$ across the parameters. The conditional distribution is distorted through biases in $\omega$, $\gamma$, and $\kappa$, which govern its scale and shape. In particular, for $\rho < 0.20$, $\omega$ is severely underestimated, while $\gamma$ is overestimated, compensating for the downward bias in $\omega$. For $\rho \geq 0.20$, we see the opposite behavior. The estimated dynamics are also affected. The autoregressive parameter $\varphi$ is underestimated for all $\rho>0$, with the largest distortion occurring around $\rho=0.25$. The score parameter $\alpha$ is overestimated for $\rho<0.20$ but becomes underestimated as the degree of heaping increases further. These results indicate that the effects of ignoring heaping are not confined to the conditional distribution itself---the misspecified model attempts to absorb features of the distribution into the dynamics of the conditional duration as well.

\begin{figure}
\centering
\includegraphics[scale=0.6]{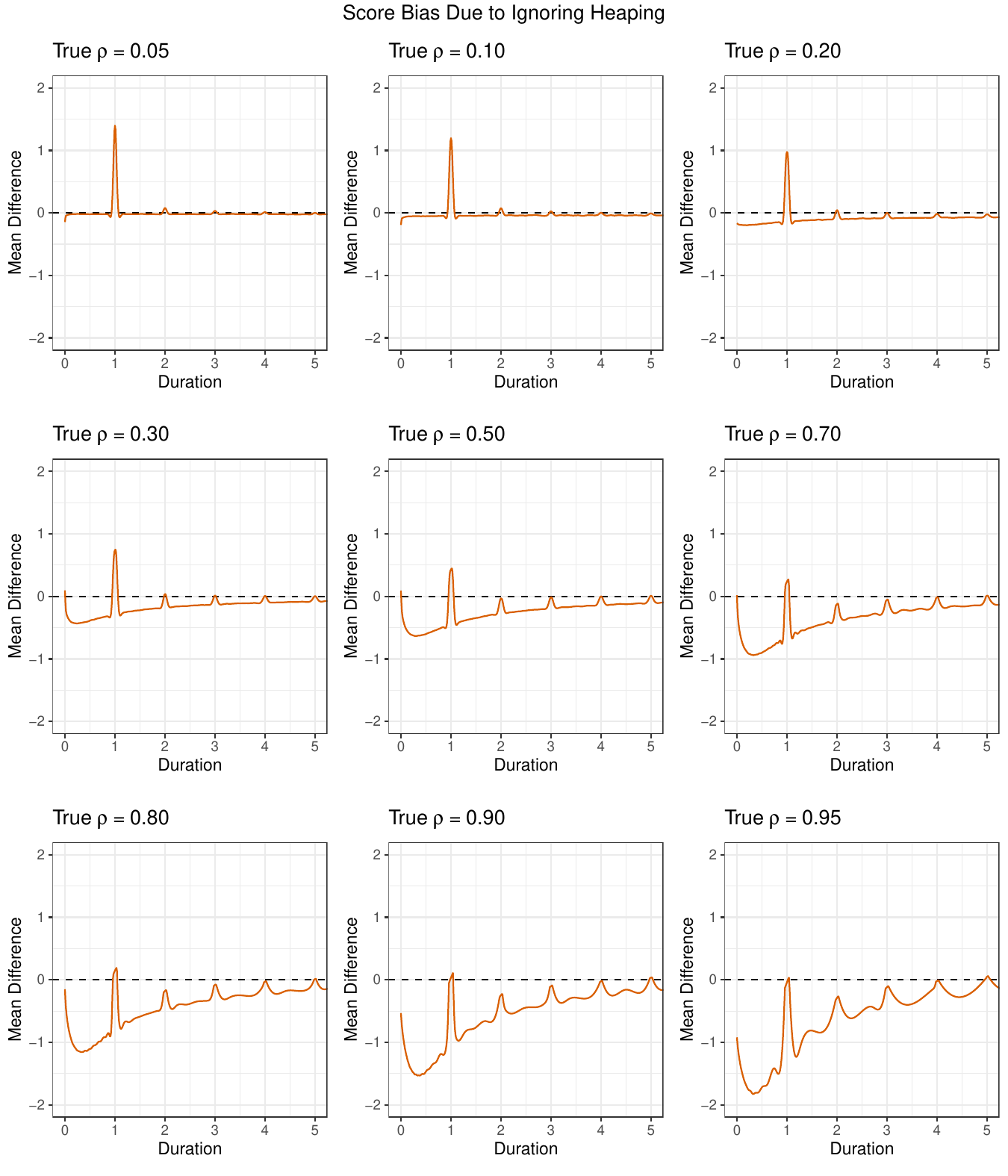}
\caption{Mean difference between the score implied by the standard ACD model and the score implied by the correctly specified GA-ACD model as a function of duration. The curve is estimated using smoothing splines. Results are based on 2,000 Monte Carlo replications.}
\label{fig:scoreBias}
\end{figure}

Figure \ref{fig:scoreBias} further shows that ignoring heaping substantially distorts the score driving the conditional dynamics. For low and moderate values of $\rho$, the score implied by the standard ACD model is markedly higher than that implied by the correctly specified GA-ACD model when the observed duration is close to an integer, particularly near one second. Between integer values, by contrast, the score difference is generally negative. As $\rho$ increases, the distortion becomes more pronounced and extends beyond the immediate neighborhoods of the integers. In particular, the score produced by the standard ACD model becomes systematically lower for smaller durations. Thus, ignoring heaping affects not only observations located at or near the heaping points but also the overall response of the conditional dynamics to observed durations.

\subsection{Performance of the GA-ACD Model}
\label{sec:simPerf}

The correctly specified GA-ACD model successfully recovers the true parameter values. Table \ref{tab:coverageMix} reports the empirical coverage probabilities of nominal 95\% confidence intervals based on the empirical Hessian. For sample sizes of $n=1\,000$ or larger, the coverage probabilities are generally close to the nominal level for all parameters except $\omega$. The confidence interval for $\omega$ exhibits substantial undercoverage, which diminishes only gradually as the sample size increases. This could be caused by highly persistent dynamics, with $\varphi = 0.998$. Overall, these results indicate that estimation and Hessian-based inference for the GA-ACD model perform well in moderately large samples, although reliable inference for $\omega$ requires considerably more observations.

\begin{table}
\centering
\caption{Empirical coverage probabilities (in percent) of 95\% confidence intervals based on the empirical Hessian for the GA-ACD model. Results are based on 2,000 Monte Carlo replications.}
\label{tab:coverageMix}
\begin{tabular}{lrrrrr}
\toprule
 & \multicolumn{5}{c}{Number of Observations} \\ \cmidrule{2-6}
 & $10^2$ & $10^3$ & $10^4$ & $10^5$ & $10^6$ \\ 
\midrule
$\omega$    & 80.13 & 79.90 & 83.00 & 90.85 & 93.95 \\ 
$\varphi$   & 84.72 & 93.15 & 93.65 & 94.80 & 95.70 \\ 
$\alpha$    & 79.68 & 94.65 & 94.85 & 94.45 & 94.95 \\ 
$\gamma$    & 79.85 & 93.90 & 94.60 & 94.65 & 94.80 \\ 
$\kappa$    & 92.81 & 94.75 & 94.95 & 94.20 & 95.00 \\ 
$\rho$      & 91.17 & 95.45 & 95.85 & 95.30 & 95.05 \\ 
$\sigma$    & 88.69 & 94.15 & 94.10 & 94.60 & 94.30 \\ 
\bottomrule
\end{tabular}
\end{table}

\section{Empirical Evidence}
\label{sec:emp}

In our empirical analysis, we focus on seven major FX pairs: EUR/USD, USD/JPY, GBP/USD, USD/CHF, AUD/USD, USD/CAD, and NZD/USD, which together represent the most actively traded currencies and capture a broad spectrum of market behavior. To provide additional context, we also examine several less frequently traded, exotic currency pairs; empirical properties for these pairs are reported in Appendix \ref{app:emp}.

\subsection{Data Sample}
\label{sec:empData}

Our dataset covers the full calendar year 2024 for each of the seven major FX pairs. Each pair contains over 10 million observations, providing a rich high-frequency sample for analysis. The data are sourced from Refinitiv Eikon. Durations are reported in seconds with three-decimal precision. A small percentage of observations are exactly zero; these are set to half the minimum increment, 0.0005 seconds, to satisfy the requirement of positive durations in the model (see, e.g., \citealp{Bauwens2006}).

\newpage

\subsection{Duration Properties}
\label{sec:empProp}

Figure \ref{fig:duration_eur} presents the empirical densities of trade durations and their fractional parts for the EUR/USD pair. The plot reveals that the distribution of durations near integer values is slightly higher immediately after the integer than just before it, suggesting a modest asymmetry in the heaping pattern. For modeling purposes, however, we adopt a symmetric truncated normal distribution to capture the spread around integers, as this simplification provides a parsimonious and tractable approach without substantially affecting the fit.

\begin{figure}
\centering
\includegraphics[scale=0.6]{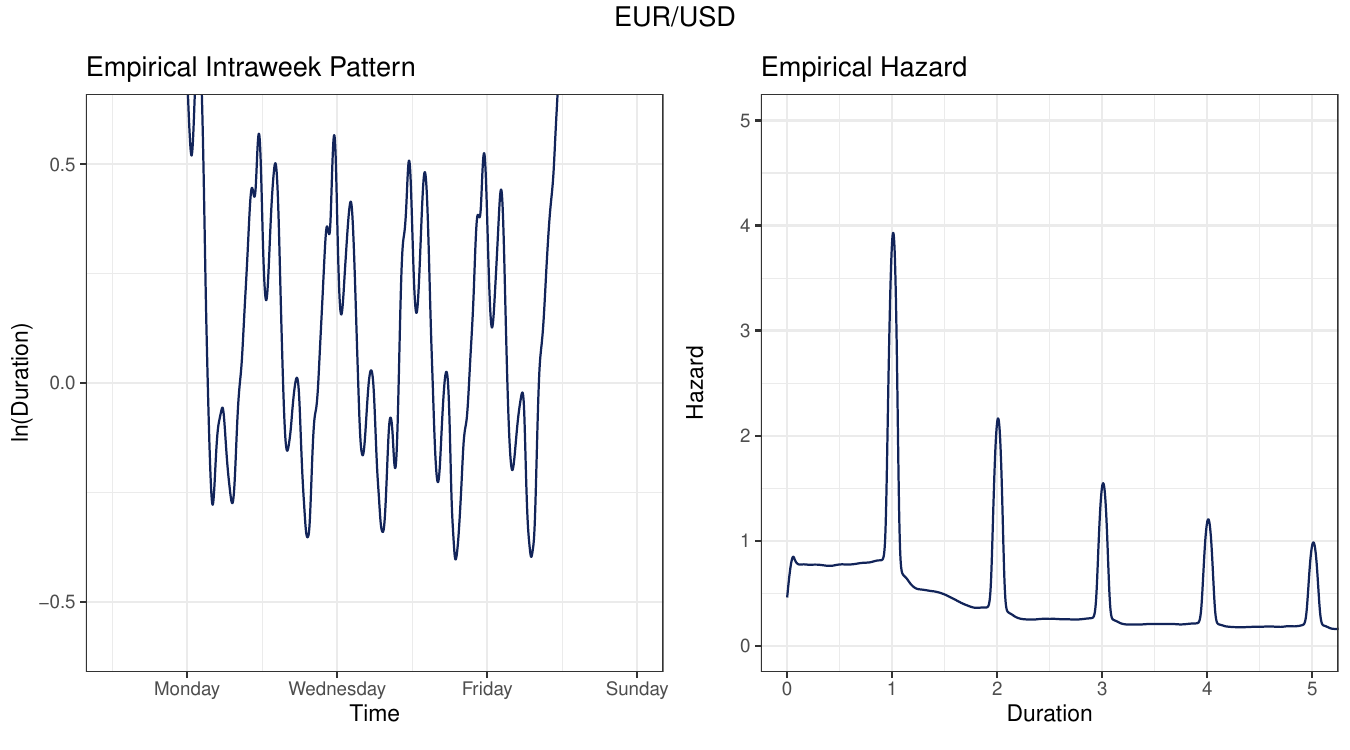}
\caption{Empirical intraweek pattern estimated with smoothing splines and empirical hazard estimated via kernel density for the EUR/USD pair.}
\label{fig:diurnal_eur}
\end{figure}

The left panel of Figure \ref{fig:diurnal_eur} illustrates the empirical intraweek pattern of trade durations for the EUR/USD pair, estimated using smoothing splines. The plot highlights clear variations in trading activity across the week, reflecting the well-known structure of the FX market. In our dataset, trading for each FX pair is recorded from 18:00 on Sunday to 22:00 on Saturday, reflecting the weekly coverage provided by Refinitiv Eikon. Activity is highest during the European and North American sessions, particularly when these sessions overlap, and considerably lower during the Asian session. These patterns reflect the global nature of the FX market and the varying liquidity and participation across time zones.

The right panel of Figure 3 shows the empirical hazard function for trade durations, estimated via a kernel density approach. The hazard is highest at very short durations, indicating clustering of trades in rapid succession, and gradually decreases for longer durations, reflecting the diminishing probability of immediate subsequent trades. In addition, the hazard exhibits pronounced spikes near integer-second durations, highlighting the heaping effect observed in the data. Together, the intraweek pattern and hazard function underscore the importance of accounting for both trading-session effects and microstructural clustering when modeling high-frequency FX durations.

\subsection{Model Fit}
\label{sec:empFit}

Tables \ref{tab:standard_major} and \ref{tab:mixture_major} report the estimated parameters and the average log-likelihood per observation for the standard score-driven ACD model based on the generalized gamma distribution and the granularity-adjusted ACD model based on the mixture distribution, respectively. We do not report standard errors, as they are extremely small in all cases due to the huge sample size, and all estimated parameters are statistically significant at any conventional significance level. Information criteria are also not reported, since the number of estimated parameters is negligible relative to the sample size, making differences in penalization terms insignificant.

\begin{figure}
\centering
\includegraphics[scale=0.6]{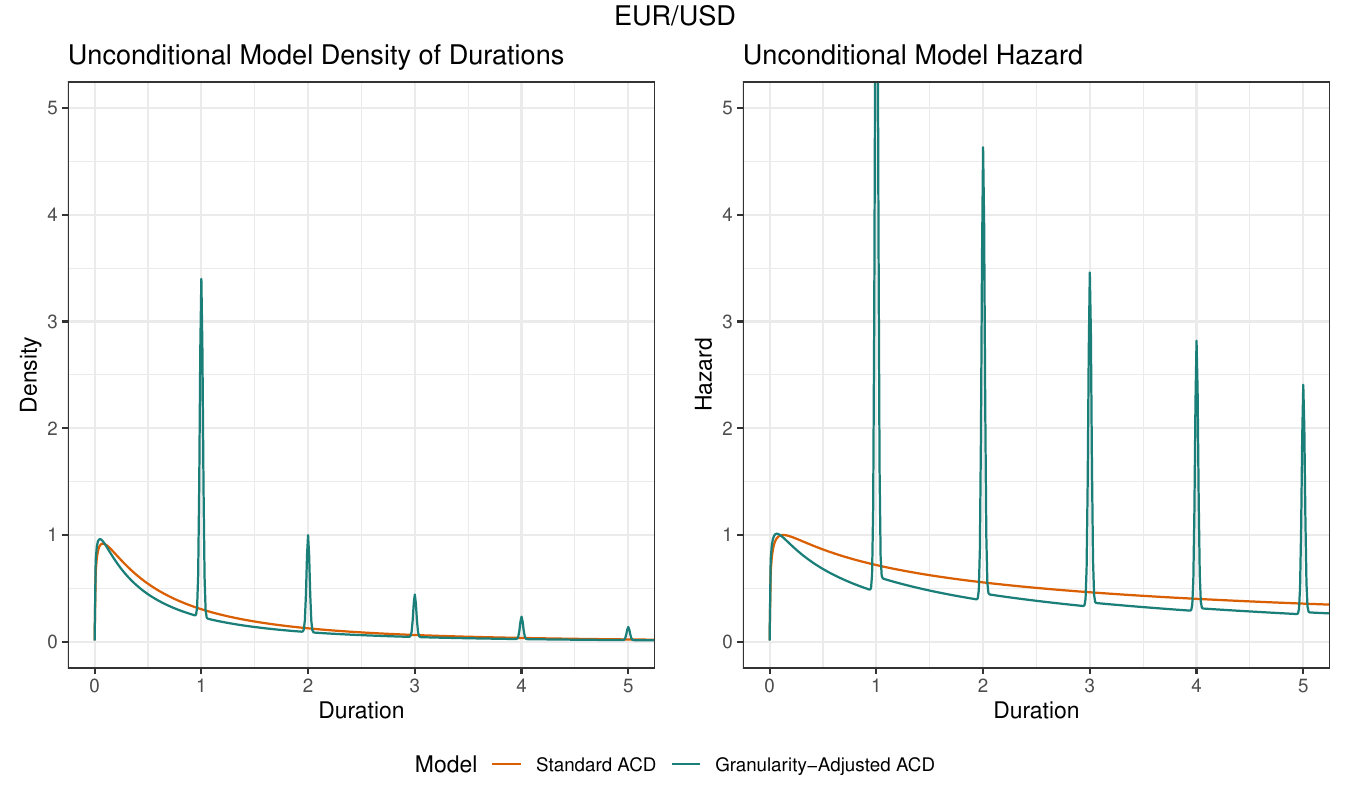}
\caption{Unconditional model density of durations and hazard for the EUR/USD pair.}
\label{fig:fit_eur}
\end{figure}

A notable difference between the two models concerns the estimated constant $\omega$, which is substantially smaller in the GA-ACD specification than in the standard ACD. This pattern is consistent with Figure \ref{fig:fit_eur}, which shows that the unconditional density implied by the standard ACD model is biased downward near integer values and upward elsewhere. In contrast, the GA-ACD model allocates more density to smaller durations---reflected in the smaller constant---while simultaneously capturing the spikes at integer values. A similar discrepancy is visible in the unconditional model hazard function displayed in Figure \ref{fig:fit_eur}.

For both specifications, the autoregressive parameter $\varphi$ is very high and close to one across all currency pairs, indicating strong duration dependence and highly persistent conditional dynamics, which is consistent with the well-documented clustering of trading activity in high-frequency financial markets. For all currency pairs, the estimated value of $\varphi$ is higher under the GA-ACD model, although the difference is relatively small for the GBP/USD and NZD/USD pairs. This finding is consistent with the results of the Monte Carlo study. The score parameter is positive in all cases, as expected, confirming that the conditional scale parameter responds appropriately to past shocks.

The estimated shape parameter $\gamma$ is higher under the GA-ACD model for most currency pairs, with the exception of the USD/CHF and NZD/USD pairs. On the other hand, the estimated value of the second shape parameter $\kappa$ is lower under the GA-ACD model for all currency pairs. These systematic differences indicate that accounting for heaping alters the estimated shape of the conditional duration distribution. Interestingly, the empirical estimates of $\gamma$ and $\kappa$, together with $\omega$, do not correspond to the simulation results at the estimated value of $\rho$, but instead resemble those obtained under higher values of $\rho$. This discrepancy may reflect differences between the true data-generating process and the GA-ACD specification.

The estimated mixture weight $\rho$ ranges from 5 to 27 percent across the seven major currency pairs, indicating that a non-negligible proportion of durations is concentrated near integer values. This provides clear quantitative evidence that clustering around integers is a systematic feature of FX trade durations. The spread parameter $\sigma$ is quite stable across the seven currency pairs, with a value of approximately 0.015.

Comparing the average log-likelihoods, the GA-ACD model delivers a clear and substantial improvement in fit across all currency pairs. The increase in average log-likelihood is sizeable given the large number of observations (over 10 million per pair), implying considerable cumulative gains in overall fit. Importantly, the superiority of the GA-ACD model is observed consistently across all major currency pairs.

\begin{table}
\centering
\begin{tabular}{lccccccc}
\toprule
& EUR/USD & USD/JPY & GBP/USD & USD/CHF & AUD/USD & USD/CAD & NZD/USD \\ 
\midrule
$\omega$  & -2.1386 & -1.0514 & -1.0959 & 0.3035 & 0.2106 & 0.2109 & 0.5461 \\ 
$\varphi$ & 0.9977 & 0.9959 & 0.9985 & 0.9978 & 0.9974 & 0.9978 & 0.9978 \\ 
$\alpha$  & 0.0315 & 0.0277 & 0.0184 & 0.0161 & 0.0224 & 0.0197 & 0.0201 \\ 
$\gamma$  & 2.9297 & 1.4398 & 1.7822 & 1.0616 & 1.1676 & 1.1778 & 1.0113 \\ 
$\kappa$  & 0.4999 & 0.7940 & 0.6915 & 0.9700 & 0.8861 & 0.8862 & 0.9876 \\ \midrule
$\ell$    & -1.2817 & -0.5016 & -0.8873 & -1.3774 & -1.4161 & -1.4461 & -1.5621 \\ 
\bottomrule
\end{tabular}
\caption{Estimated parameters with average log-likelihood $\ell$ for the standard ACD model based on the generalized gamma distribution across the seven major pairs.}
\label{tab:standard_major}
\end{table}

\begin{table}
\centering
\begin{tabular}{lccccccc}
\toprule
& EUR/USD & USD/JPY & GBP/USD & USD/CHF & AUD/USD & USD/CAD & NZD/USD \\ 
\midrule
$\omega$  & -3.3459 & -1.2583 & -1.4993 & 0.1539 & -0.0321 & -0.0762 & 0.4482 \\ 
$\varphi$ & 0.9985 & 0.9969 & 0.9985 & 0.9981 & 0.9975 & 0.9980 & 0.9978 \\ 
$\alpha$  & 0.0280 & 0.0246 & 0.0193 & 0.0208 & 0.0277 & 0.0247 & 0.0252 \\ 
$\gamma$  & 3.7112 & 1.5542 & 2.0269 & 1.0547 & 1.2236 & 1.2214 & 0.9740 \\ 
$\kappa$  & 0.4142 & 0.7522 & 0.6261 & 0.8891 & 0.7936 & 0.7657 & 0.9086 \\ 
$\rho$    & 0.1861 & 0.0501 & 0.0837 & 0.2055 & 0.1892 & 0.2734 & 0.2324 \\ 
$\sigma$  & 0.0146 & 0.0148 & 0.0165 & 0.0145 & 0.0145 & 0.0153 & 0.0150 \\ \midrule
$\ell$    & -1.0465 & -0.4653 & -0.8209 & -1.1142 & -1.1852 & -1.0455 & -1.2592 \\ 
\bottomrule
\end{tabular}
\caption{Estimated parameters with average log-likelihood $\ell$ for the granularity-adjusted ACD model based on the mixture distribution across the seven major pairs.}
\label{tab:mixture_major}
\end{table}

\subsection{Note on Computation}
\label{sec:empComp}

The GA-ACD model was estimated in R using the gasmodel package developed by \cite{Holy2026}, modified to incorporate the proposed mixture distribution. Model estimation becomes computationally intensive for large datasets, particularly those containing more than 10 million observations, owing to the complex form of the score function. Replacing the score of the proposed mixture distribution with the simpler score of the generalized gamma distribution considerably reduces the computational burden, although this computational gain comes at the cost of a poorer empirical fit.

\section{Conclusion}
\label{sec:con}

In this paper, we propose the granularity-adjusted autoregressive conditional duration (GA-ACD) model to account for the heaping of trade durations at and around integer seconds observed in high-frequency FX data. The model combines a standard generalized gamma component for regular durations with a second component that locally redistributes probability mass around integer values, while the conditional dynamics are specified within a score-driven framework. The simulation results show that ignoring heaping leads to systematic distortions in the estimated distribution and dynamics, whereas the GA-ACD model reliably recovers the underlying parameters. Empirically, the GA-ACD model provides a better fit than the standard generalized gamma ACD model across the major currency pairs considered, demonstrating the importance of explicitly accounting for heaping in FX trade durations.

For future research, the proposed framework can be extended to accommodate alternative baseline distributions beyond the generalized gamma distribution, thereby allowing greater flexibility in modeling duration distributions and dynamics. A similar mixture-based approach could also be applied to exotic currency pairs; although these pairs exhibit more complex and heterogeneous clustering patterns that may require additional modeling refinements.

\section*{Funding}
\label{sec:fund}

The work on this paper was supported by the Czech Science Foundation under project 26-22844S and the personal and professional development support program of the Faculty of Informatics and Statistics, Prague University of Economics and Business.

\section*{Use of Artificial Intelligence}
\label{sec:ai}

Generative artificial intelligence tools were used to improve the grammar, clarity, and style of the manuscript. The author reviewed and revised all AI-assisted text and take full responsibility for the content of the paper.


\appendix

\section{Details of the Mixture Distribution}
\label{app:model}

The cumulative distribution function of the mixture distribution is given by
\begin{equation}
F_{X_i}(x_i) = (1 - \rho) F_{Y_i}(x_i) + \rho F_{Z_i}(x_i).
\end{equation}
The cumulative distribution function of the generalized gamma distribution is given by
\begin{equation}
F_{Y_i}(x_i) = P \left( \gamma, \left( \frac{x_i}{\lambda_i} \right)^\kappa \right), 
\end{equation}
where $P(\cdot,\cdot)$ is the regularized lower incomplete gamma function. The cumulative distribution function of the distribution concentrated near integers is given by
\begin{equation}
F_{Z_i}(x_i) = \frac{ \left( F_{Y_i}(\lfloor x_i \rceil) - F_{Y_i}(\lfloor x_i \rceil - 1) \right) \left( F_{W_i}(x_i) - F_{W_i}(\lfloor x_i \rceil - 0.5) \right) }{ F_{W_i}(\lfloor x_i \rceil + 0.5) - F_{W_i}(\lfloor x_i \rceil - 0.5) } + F_{Y_i}(\lfloor x_i \rceil - 1).
\end{equation}
The score of the mixture distribution with respect to $\ln \lambda_i$ is given by
\begin{equation}
\nabla_{X_i; \ln \lambda_i}(x_i) = (1 - \rho) \frac{f_{Y_i}(x_i)}{f_{X_i}(x_i)} \nabla_{Y_i; \ln \lambda_i}(x_i) + \rho \frac{f_{Z_i}(x_i)}{f_{X_i}(x_i)} \nabla_{Z_i; \ln \lambda_i}(x_i).
\end{equation}
The score of the generalized gamma distribution with respect to $\ln \lambda_i$ is given by
\begin{equation}
\nabla_{Y_i; \ln \lambda_i}(x_i) = \kappa \left( \left( \frac{x_i}{\lambda_i} \right)^{\kappa} - \gamma \right).
\end{equation}
The score of the distribution concentrated near integers with respect to $\ln \lambda_i$ is given by
\begin{equation}
\begin{aligned}
\nabla_{Z_i; \ln \lambda_i}(x_i) &= \frac{1}{F_{Y_i}(\lfloor x_i \rceil) - F_{Y_i}(\lfloor x_i \rceil - 1)} \left( \frac{\partial \ln F_{Y_i}(\lfloor x_i \rceil)}{\partial \ln \lambda_i} - \frac{\partial \ln F_{Y_i}(\lfloor x_i \rceil - 1)}{\partial \ln \lambda_i} \right) \\
&= \frac{\kappa}{\Gamma(\gamma)} \frac{ \left( \frac{\lfloor x_i \rceil - 1}{\lambda_i} \right)^{\gamma \kappa} \exp \left( - \left( \frac{\lfloor x_i \rceil - 1}{\lambda_i} \right)^{\kappa} \right) - \left( \frac{\lfloor x_i \rceil}{\lambda_i} \right)^{\gamma \kappa} \exp \left( - \left( \frac{\lfloor x_i \rceil}{\lambda_i} \right)^{\kappa} \right) }{P \left( \gamma, \left( \frac{\lfloor x_i \rceil}{\lambda_i} \right)^\kappa \right) - P \left( \gamma, \left( \frac{\lfloor x_i \rceil - 1}{\lambda_i} \right)^\kappa \right)}.
\end{aligned}
\end{equation}
The distribution concentrated near integers can be expressed as an infinite mixture:
\begin{equation}
Z_i = \begin{cases}
V_{i,1} & \text{with probability } F_{Y_i}(1), \\
V_{i,2} & \text{with probability } F_{Y_i}(2) - F_{Y_i}(1), \\
V_{i,3} & \text{with probability } F_{Y_i}(3) - F_{Y_i}(2), \\
\ \vdots \\
\end{cases}
\end{equation}
where $V_{i,k}$ follow the normal distribution with mean $k$ and standard deviation $\sigma$ truncated to $(k - 0.5, k + 0.5)$. The moments of the truncated normal distribution are given by
\begin{equation}
\begin{aligned}
\mathrm{E}[V_{i,k}] &= k, \\
\mathrm{var}[V_{i,k}] &= \sigma^2 \left( 1 - \frac{\frac{1}{\sigma} \varphi \left( \frac{0.5}{\sigma} \right) }{2 \Phi \left( \frac{0.5}{\sigma} \right) - 1} \right), \\
\end{aligned}
\end{equation}
where $\varphi(\cdot)$ and $\Phi(\cdot)$ are the density and cumulative distribution functions of the standard normal distribution. Note that $\mathrm{var}[V_{i,k}]$ is the same for all $k$. The expected value of the mixture distribution is given by
\begin{equation}
\mathrm{E}[X_i] = (1 - \rho) \mathrm{E}[Y_i ] + \rho \mathrm{E}[Z_i].
\end{equation}
The expected value of the generalized gamma distribution is given by
\begin{equation}
\mathrm{E}[Y_i] = \lambda_i \frac{\Gamma \left( \gamma + \frac{1}{\kappa} \right)}{\Gamma \left( \gamma \right)}.
\end{equation}
The expected value of the distribution concentrated near integers is given by
\begin{equation}
\begin{aligned}
\mathrm{E}[Z_i] &= \sum_{k=1}^{\infty} \left( F_{Y_i}(k) - F_{Y_i}(k - 1) \right) \mathrm{E}[V_{i,k}] \\
&= \sum_{k=1}^{\infty} \left( F_{Y_i}(k) - F_{Y_i}(k - 1) \right) k \\
&= \sum_{k=1}^{\infty} \left( 1 - F_{Y_i}(k - 1) \right). \\
\end{aligned}
\end{equation}
In the special case of the exponential distribution ($\gamma = \kappa = 1$), this has the closed form
\begin{equation}
\mathrm{E}[Z_i] = \frac{1}{1 - \exp \left( - \frac{1}{\lambda_i} \right)}.
\end{equation}
The variance of the mixture distribution is given by
\begin{equation}
\mathrm{var}[X_i] = (1 - \rho) \left( \mathrm{var}[Y_i] + \mathrm{E}[Y_i ]^2 \right) + \rho \left( \mathrm{var}[Z_i] + \mathrm{E}[Z_i ]^2 \right) - \mathrm{E}[X_i ]^2.
\end{equation}
The variance of the generalized gamma distribution is given by
\begin{equation}
\mathrm{var}[Y_i] = \lambda_i^2 \left( \frac{\Gamma \left(\gamma + \frac{2}{\kappa} \right)}{\Gamma \left( \gamma \right)} -  \left( \frac{\Gamma \left(\gamma + \frac{1}{\kappa} \right)}{\Gamma \left( \gamma \right)} \right)^2 \right).
\end{equation}
The variance of the distribution concentrated near integers is given by
\begin{equation}
\begin{aligned}
\mathrm{var}[Z_i] &= \sum_{k=1}^\infty \left( F_{Y_i}(k) - F_{Y_i}(k - 1) \right) \left( \mathrm{var}[V_{i,k}] + \mathrm{E}[V_{i,k}]^2 \right) - \mathrm{E}[Z_i]^2 \\
&= \mathrm{var}[V_{i,1}] + \sum_{k=1}^\infty \left( F_{Y_i}(k) - F_{Y_i}(k - 1) \right) k^2 - \mathrm{E}[Z_i]^2 \\
&= \mathrm{var}[V_{i,1}] + \sum_{k=1}^{\infty} (2k - 1) \left( 1 - F_{Y_i}(k - 1) \right) - \mathrm{E}[Z_i]^2. \\
\end{aligned}
\end{equation}
In the special case of the exponential distribution ($\gamma = \kappa = 1$), this has the closed form
\begin{equation}
\mathrm{var}[Z_i] = \mathrm{var}[V_{i,1}] + \frac{\exp \left( -\frac{1}{\lambda_i} \right)}{\left( 1 - \exp \left( -\frac{1}{\lambda_i} \right) \right)^2}.
\end{equation}

\section{Further Simulation Results}
\label{app:sim}

\begin{figure}
\centering
\includegraphics[scale=0.6]{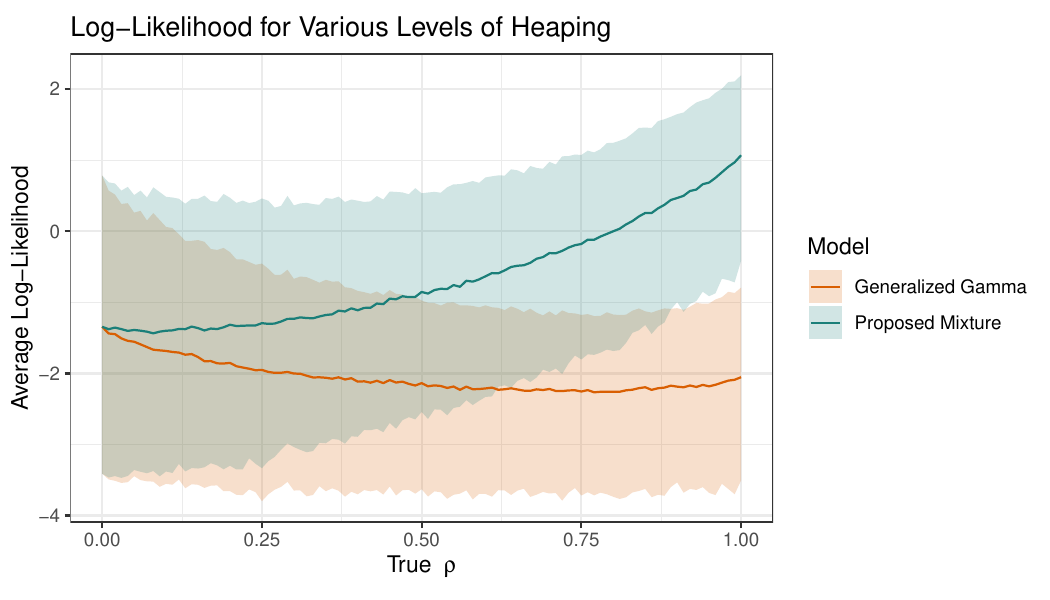}
\caption{Average log-likelihood together with the 95\%t interval across simulation replications for the standard ACD and GA-ACD models. Results are based on 2,000 Monte Carlo replications.}
\label{fig:loglik}
\end{figure}

\begin{figure}
\centering
\includegraphics[scale=0.6]{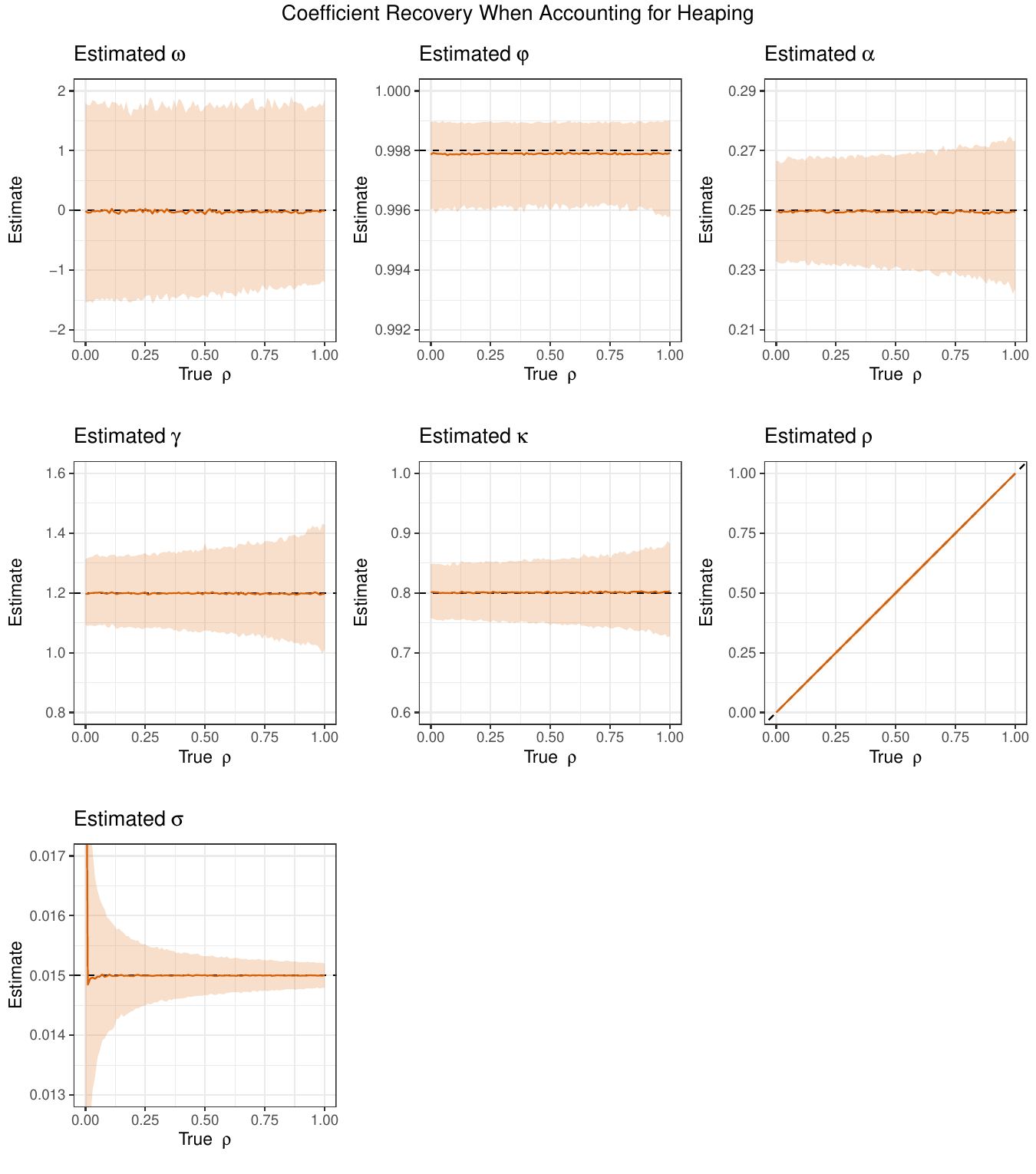}
\caption{Median parameter estimates together with the 95\% interval across simulation replications for the GA-ACD model. Results are based on 2,000 Monte Carlo replications.}
\label{fig:coefRec}
\end{figure}

\begin{figure}
\centering
\includegraphics[scale=0.6]{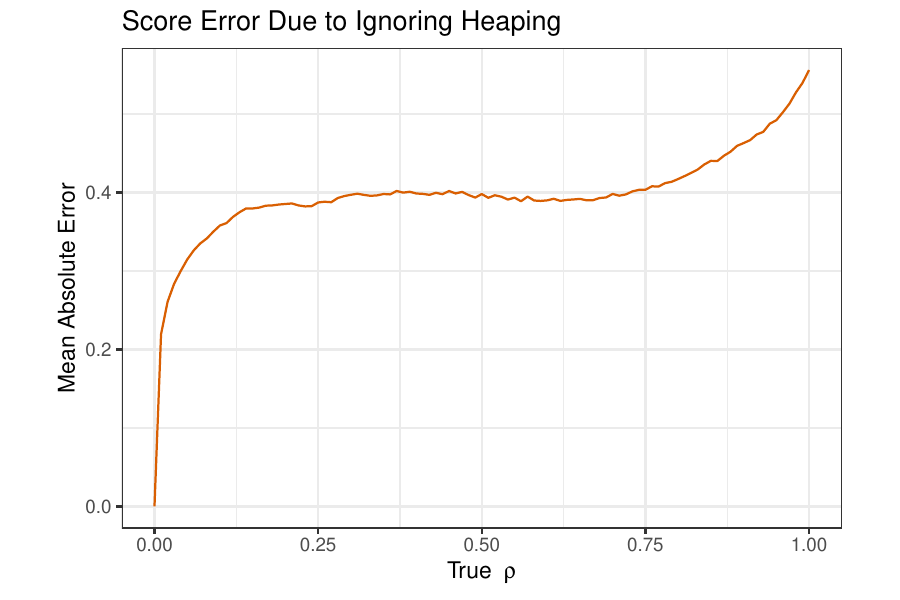}
\caption{Mean absolute difference between the score implied by the standard ACD model and the score implied by the correctly specified GA-ACD model as a function of the mixture weight $\rho$. Results are based on 2,000 Monte Carlo replications.}
\label{fig:scoreError}
\end{figure}

This appendix reports additional results from the Monte Carlo study. Figure \ref{fig:loglik} shows that the log-likelihood of the misspecified standard ACD model decreases as $\rho$ increases, reflecting the model's inability to capture increasingly pronounced heaping; by contrast, the log-likelihood of the correctly specified GA-ACD model increases as the model adequately captures the growing concentration of observations around integer values. Figure \ref{fig:coefRec} demonstrates that the correctly specified GA-ACD model accurately recovers the true parameter values. Figure \ref{fig:scoreError} shows that the absolute difference between the scores implied by the misspecified standard ACD model and the correctly specified GA-ACD model increases with $\rho$. 

\section{Further Empirical Evidence}
\label{app:emp}

\begin{figure}
\centering
\includegraphics[scale=0.6]{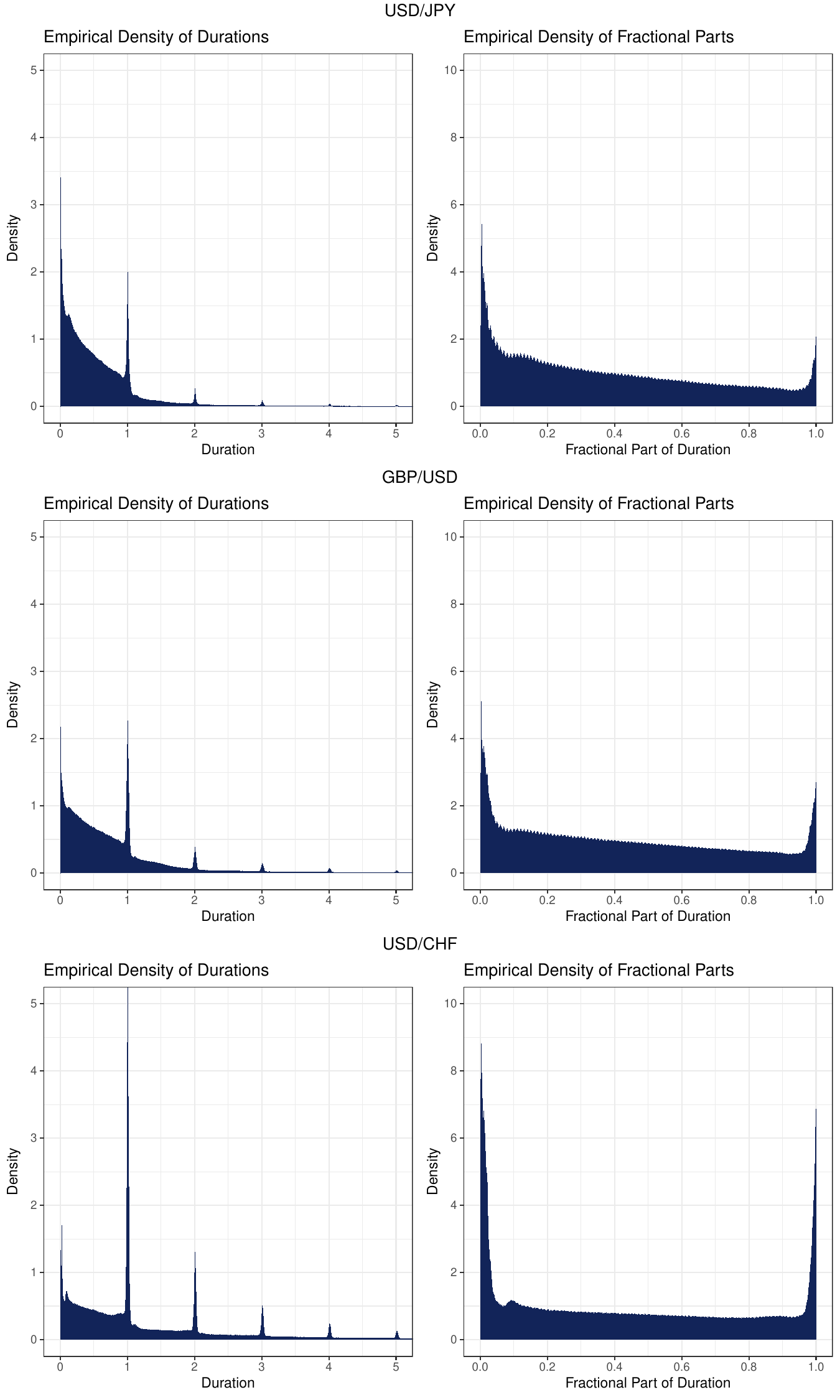}
\caption{Empirical densities of trade durations and their fractional part for the USD/JPY, GBP/USD, and USD/CHF pairs.}
\label{fig:duration_major1}
\end{figure}

\begin{figure}
\centering
\includegraphics[scale=0.6]{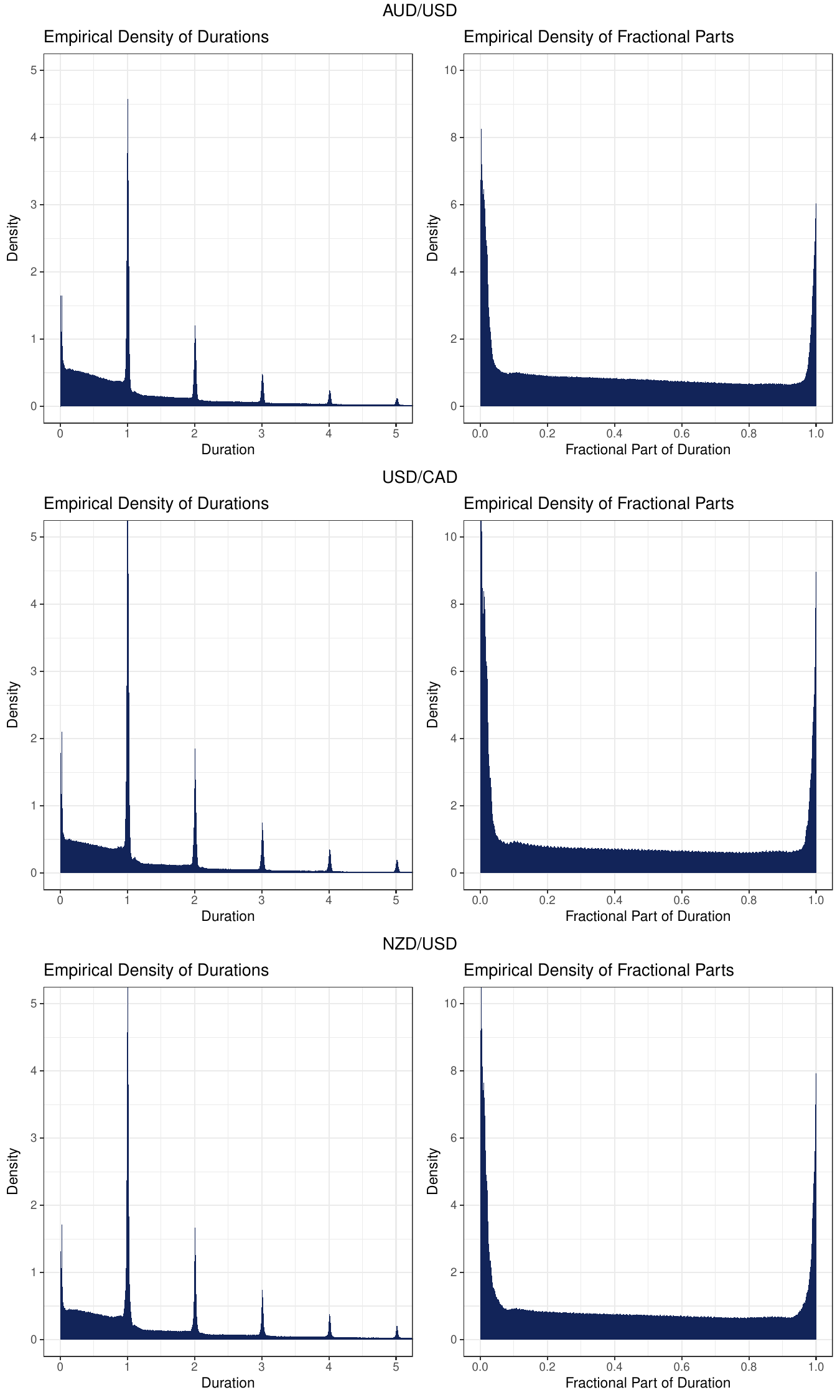}
\caption{Empirical densities of trade durations and their fractional part for the AUD/USD, USD/CAD, and NZD/USD pairs.}
\label{fig:duration_major2}
\end{figure}

\begin{figure}
\centering
\includegraphics[scale=0.6]{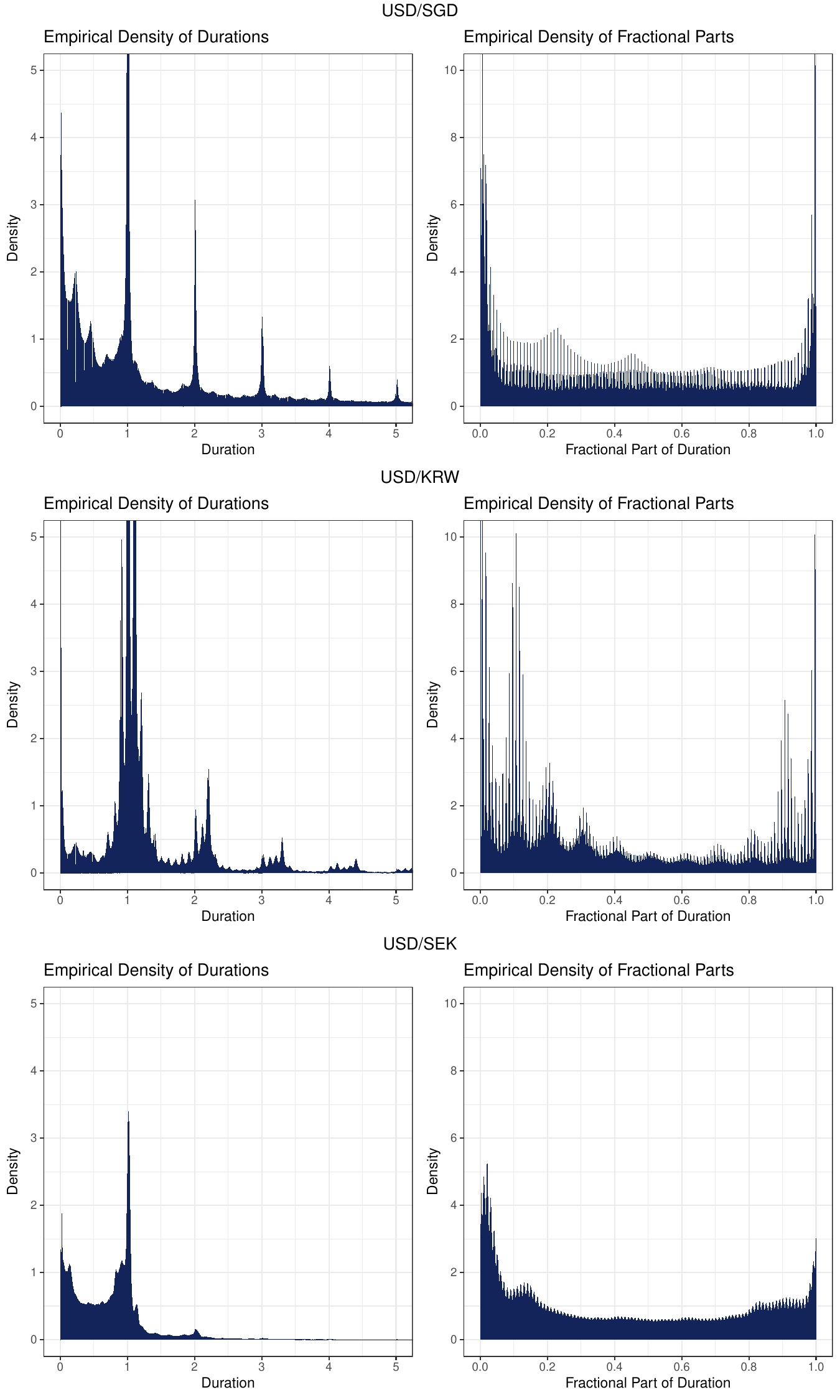}
\caption{Empirical densities of trade durations and their fractional part for the USD/SGD, USD/KRW, and USD/SEK pairs.}
\label{fig:duration_exotic1}
\end{figure}

\begin{figure}
\centering
\includegraphics[scale=0.6]{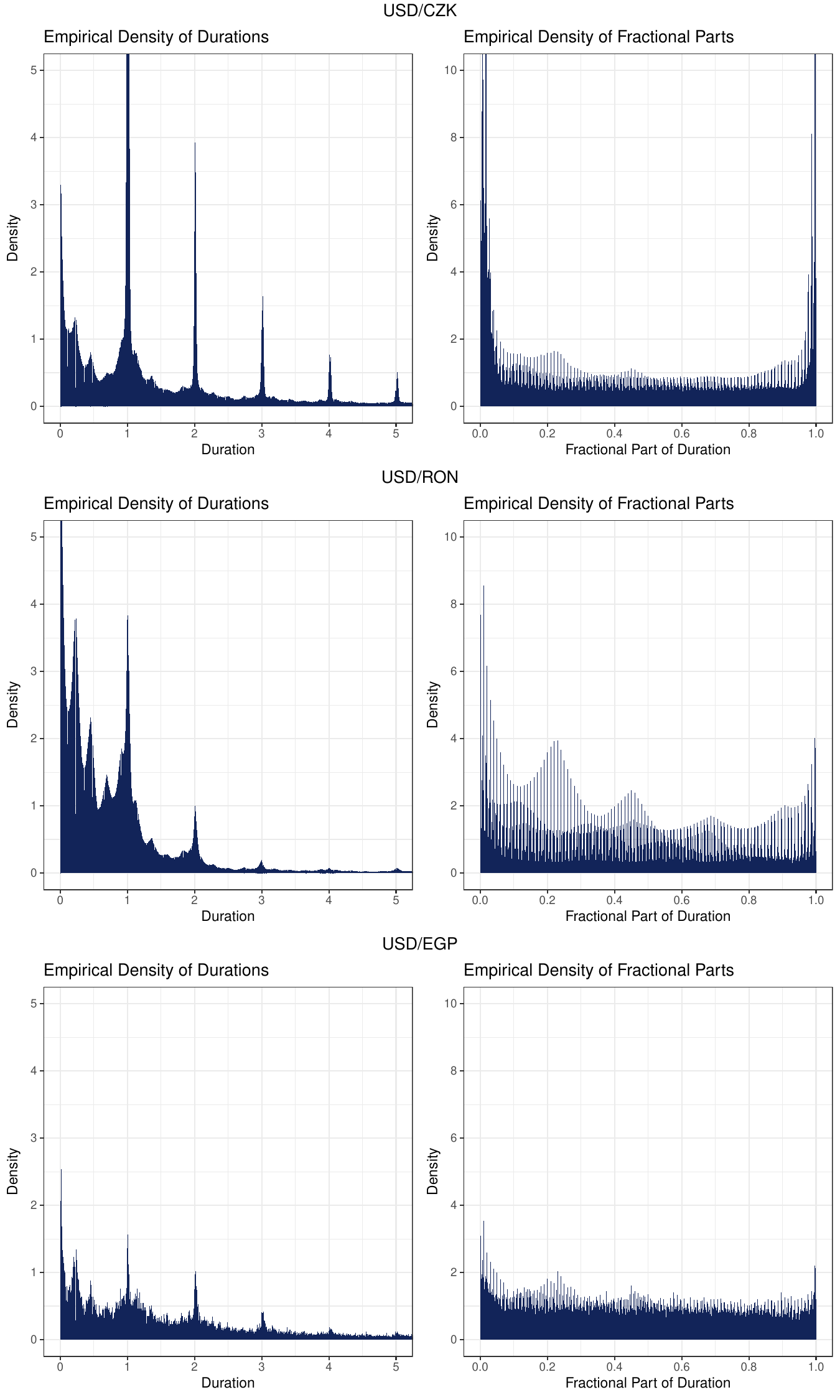}
\caption{Empirical densities of trade durations and their fractional part for the USD/CZK, USD/RON, and USD/EGP pairs.}
\label{fig:duration_exotic2}
\end{figure}

\begin{figure}
\centering
\includegraphics[scale=0.6]{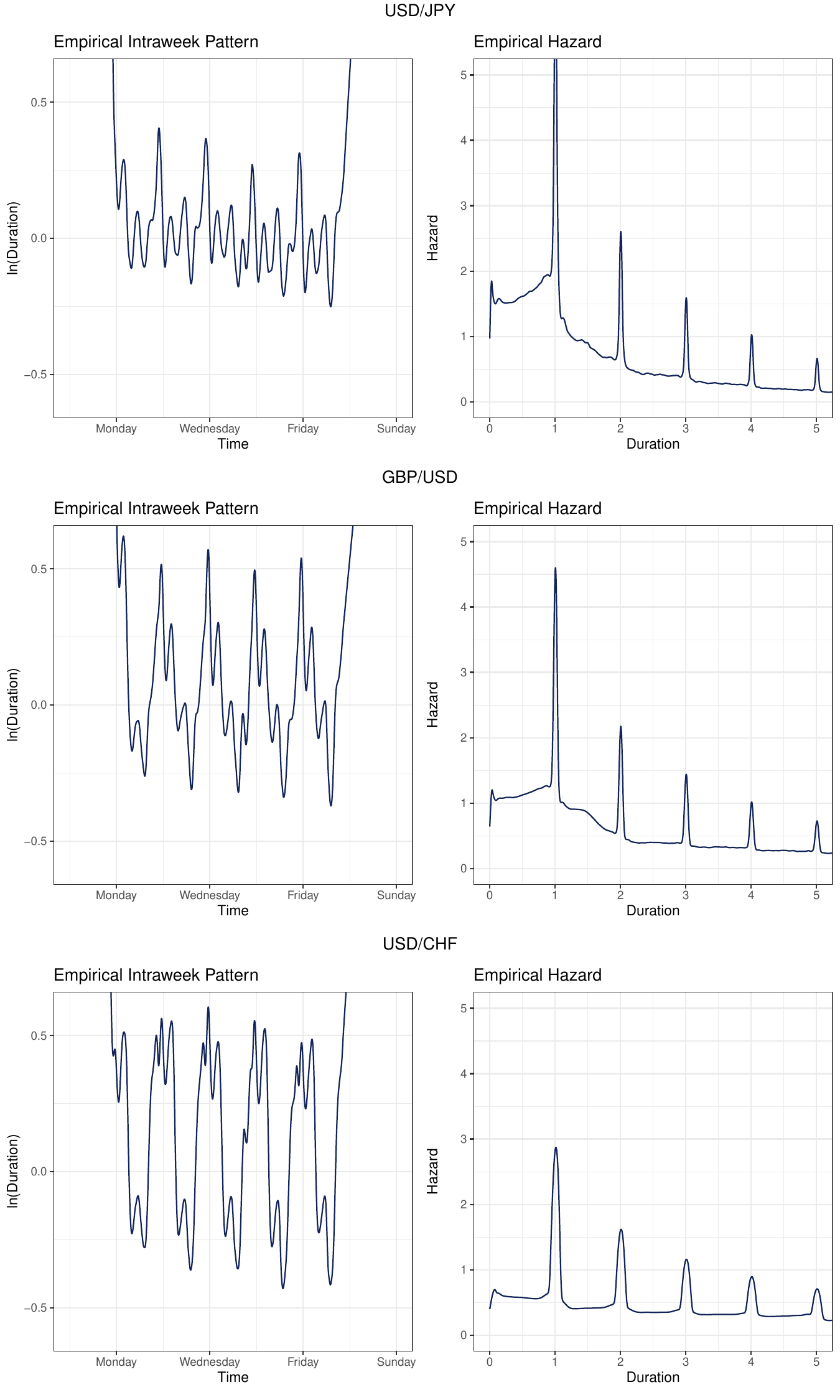}
\caption{Empirical intraweek pattern estimated with smoothing splines and empirical hazard estimated via kernel density for the USD/JPY, GBP/USD, and USD/CHF pairs.}
\label{fig:diurnal_major1}
\end{figure}

\begin{figure}
\centering
\includegraphics[scale=0.6]{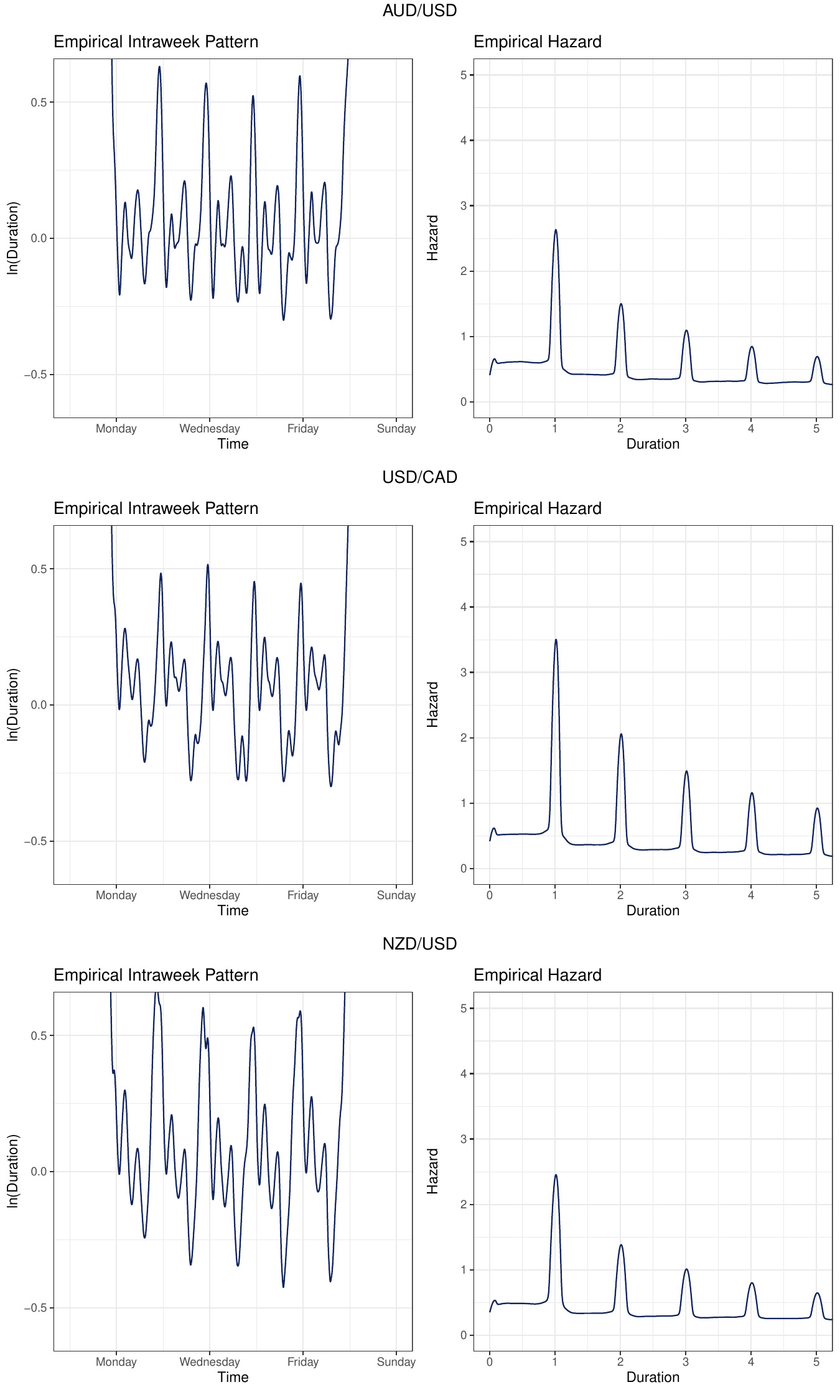}
\caption{Empirical intraweek pattern estimated with smoothing splines and empirical hazard estimated via kernel density for the AUD/USD, USD/CAD, and NZD/USD pairs.}
\label{fig:diurnal_major2}
\end{figure}

\begin{figure}
\centering
\includegraphics[scale=0.6]{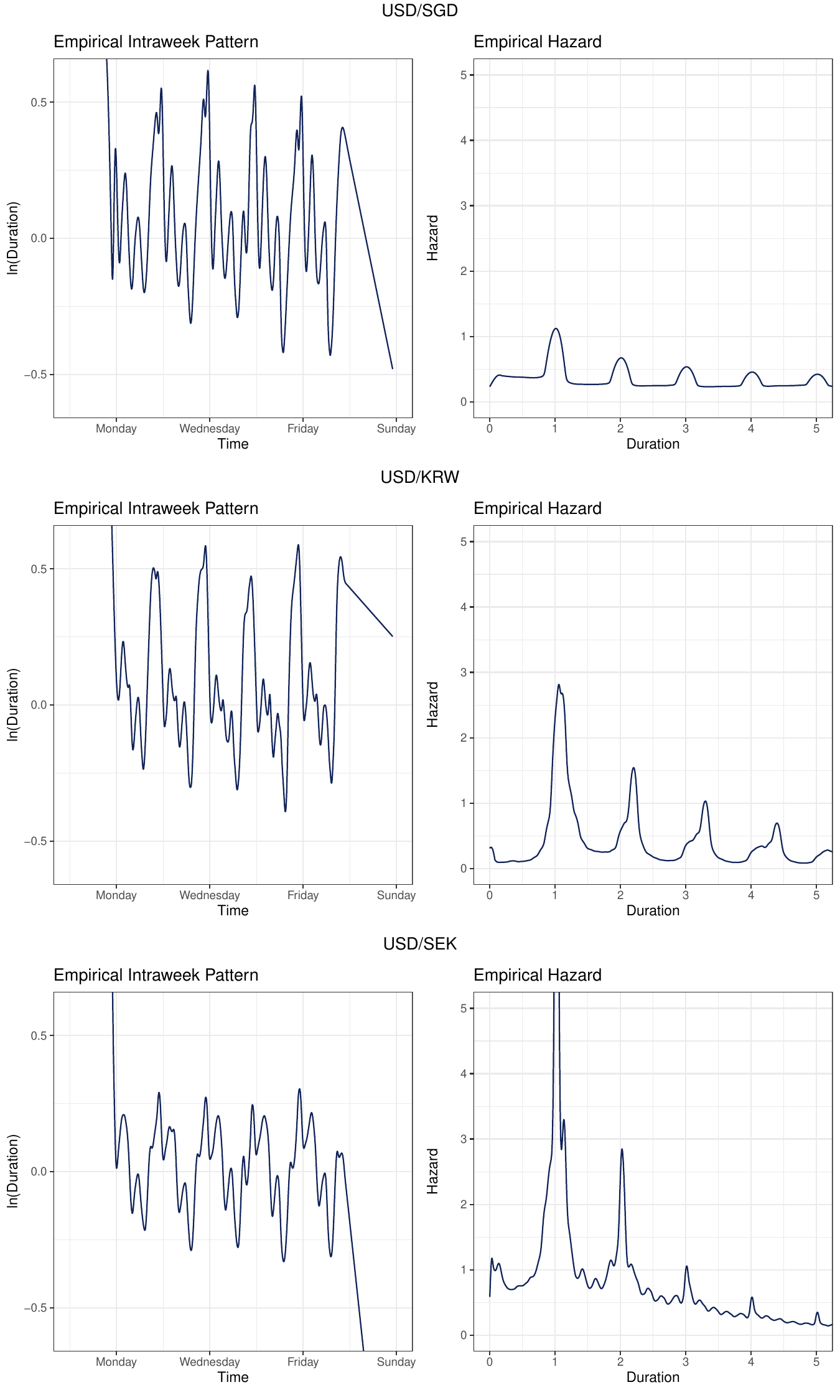}
\caption{Empirical intraweek pattern estimated with smoothing splines and empirical hazard estimated via kernel density for the USD/SGD, USD/KRW, and USD/SEK pairs.}
\label{fig:diurnal_exotic1}
\end{figure}

\begin{figure}
\centering
\includegraphics[scale=0.6]{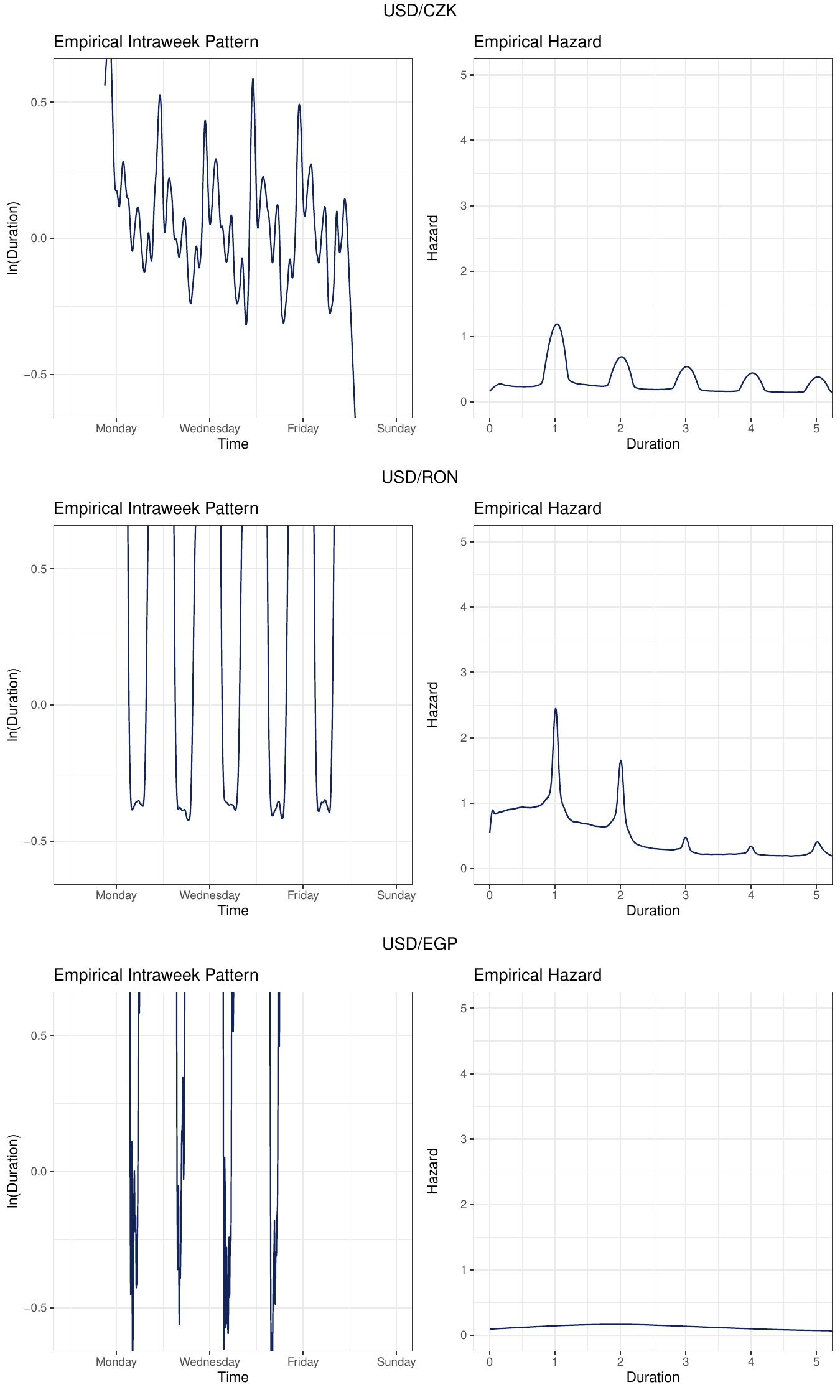}
\caption{Empirical intraweek pattern estimated with smoothing splines and empirical hazard estimated via kernel density for the USD/CZK, USD/RON, and USD/EGP pairs.}
\label{fig:diurnal_exotic2}
\end{figure}

\begin{figure}
\centering
\includegraphics[scale=0.6]{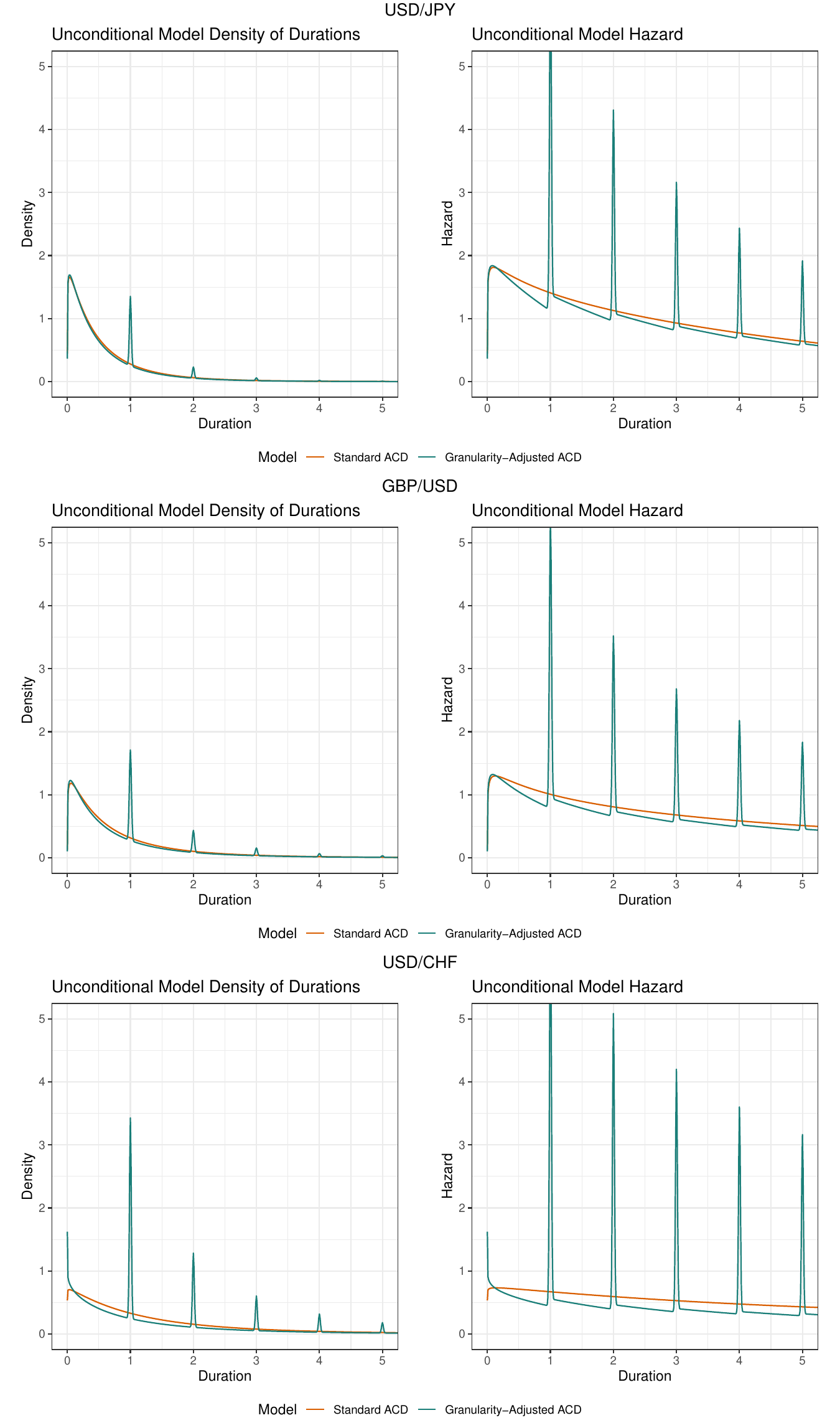}
\caption{Unconditional model density of durations and hazard for the USD/JPY, GBP/USD, and USD/CHF pairs.}
\label{fig:fit_major1}
\end{figure}

\begin{figure}
\centering
\includegraphics[scale=0.6]{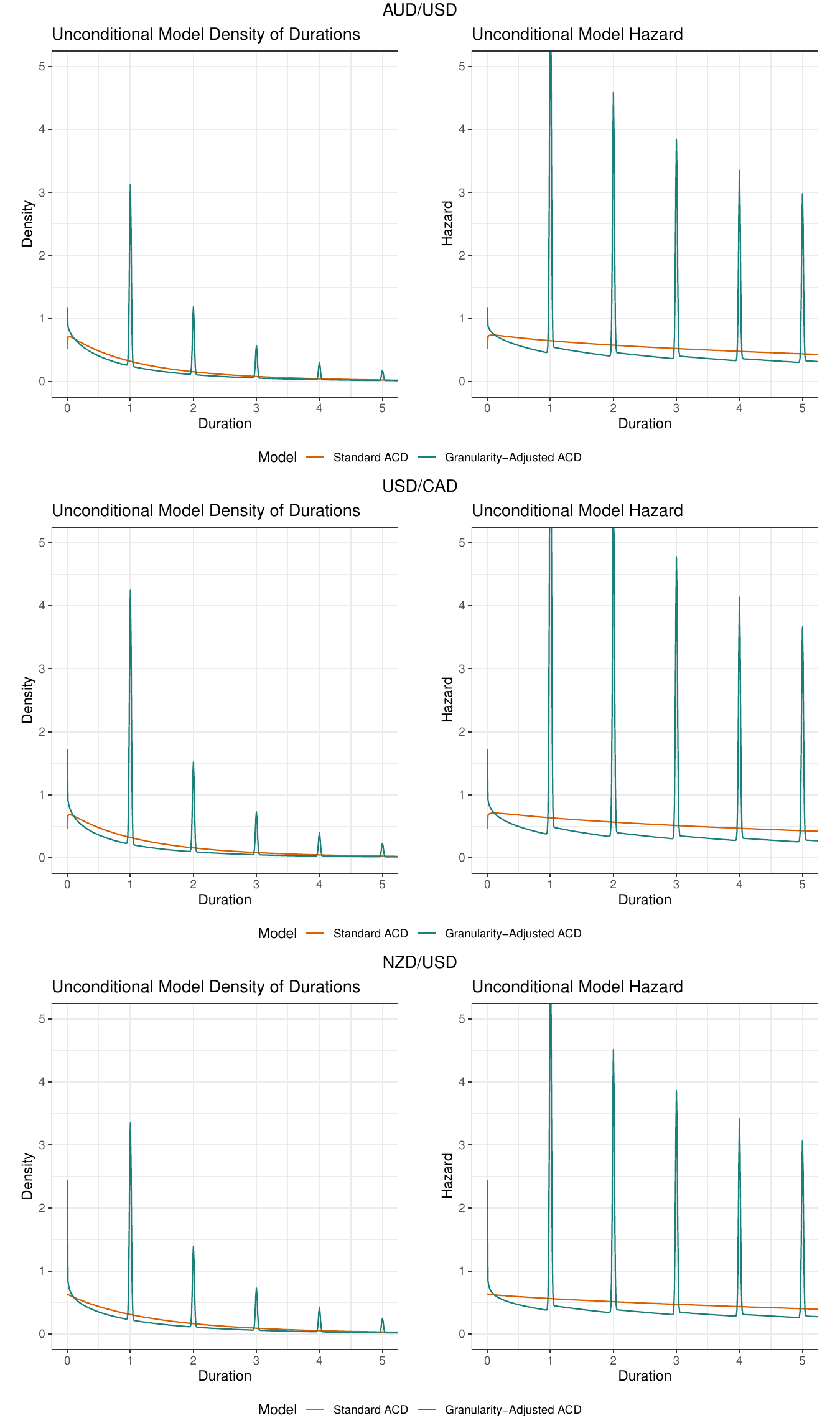}
\caption{Unconditional model density of durations and hazard for the AUD/USD, USD/CAD, and NZD/USD pairs.}
\label{fig:fit_major2}
\end{figure}

\begin{figure}
\centering
\includegraphics[scale=0.6]{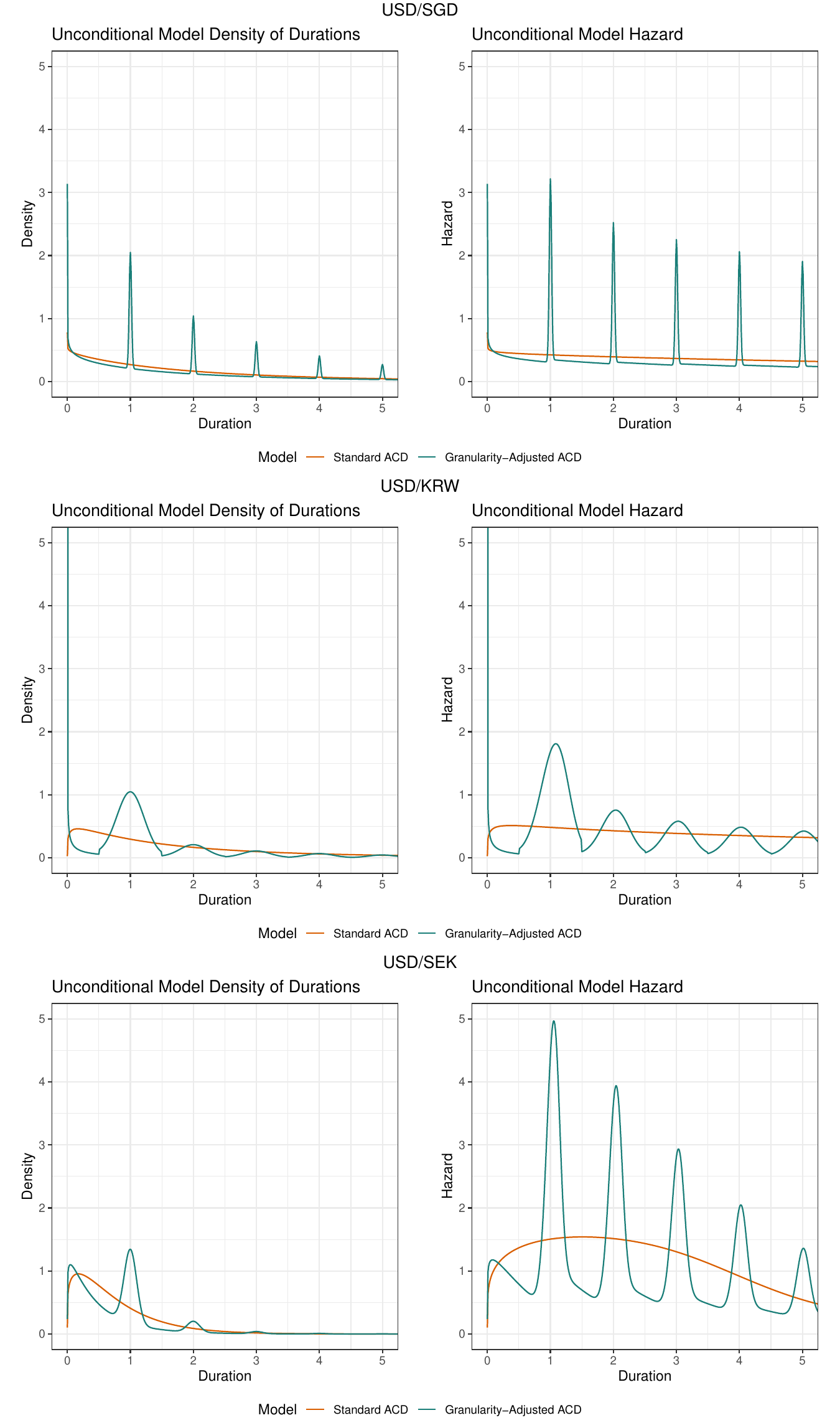}
\caption{Unconditional model density of durations and hazard for the USD/SGD, USD/KRW, and USD/SEK pairs.}
\label{fig:fit_exotic1}
\end{figure}

\begin{figure}
\centering
\includegraphics[scale=0.6]{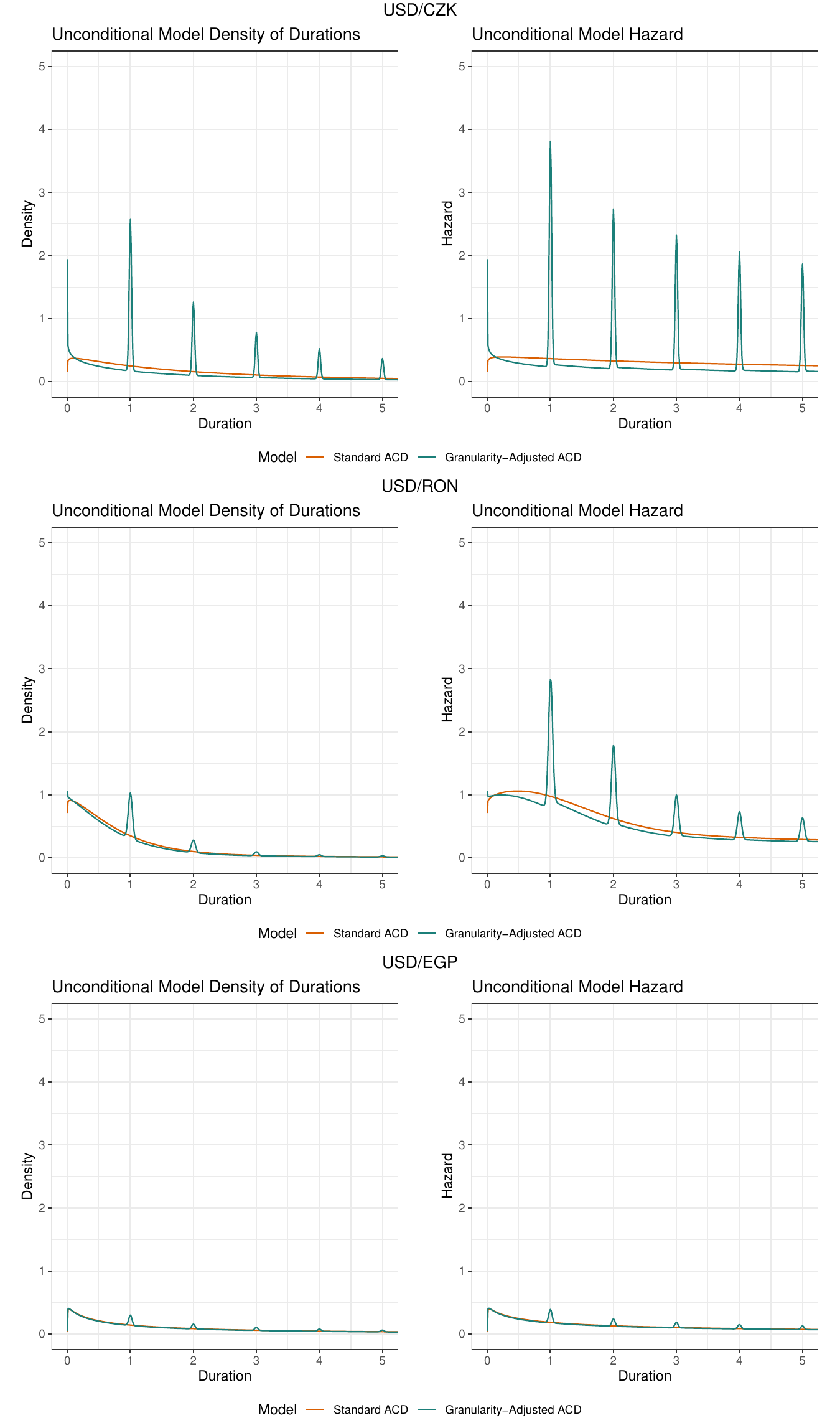}
\caption{Unconditional model density of durations and hazard for the USD/CZK, USD/RON, and USD/EGP pairs.}
\label{fig:fit_exotic2}
\end{figure}

This appendix reports the results for the remaining six major pairs and six selected exotic pairs, namely USD/SGD, USD/KRW, USD/SEK, USD/CZK, USD/RON, and USD/EGP. Figures \ref{fig:duration_major1}, \ref{fig:duration_major2}, \ref{fig:duration_exotic1}, and \ref{fig:duration_exotic2} show the empirical densities of durations and their fractional part. Figures \ref{fig:diurnal_major1}, \ref{fig:diurnal_major2}, \ref{fig:diurnal_exotic1}, and \ref{fig:diurnal_exotic2} show the empirical intraweek pattern and empirical hazard. Figures \ref{fig:fit_major1}, \ref{fig:fit_major2}, \ref{fig:fit_exotic1}, and \ref{fig:fit_exotic2} show the unconditional model density of durations and hazard. Finally, Tables \ref{tab:standard_exotic} and \ref{tab:mixture_exotic} report estimated parameters.

\begin{table}
\centering
\begin{tabular}{lcccccc}
\toprule
& USD/SGD & USD/KRW & USD/SEK & USD/CZK & USD/RON & USD/EGP \\
\midrule
$\omega$  & 0.9517 & -0.2805 & -0.1729 & 0.7461 & 0.2690 & -2.0345 \\ 
$\varphi$ & 0.9954 & 0.9890 & 0.9917 & 0.9978 & 0.9991 & 0.9701 \\ 
$\alpha$  & 0.0318 & 0.0501 & 0.0153 & 0.0318 & 0.0048 & 0.4362 \\ 
$\gamma$  & 0.9256 & 1.8825 & 0.9580 & 1.2575 & 0.7801 & 3.9198 \\ 
$\kappa$  & 1.0344 & 0.6756 & 1.2492 & 0.8628 & 1.3138 & 0.3412 \\ \midrule
$\ell$    & -1.8618 & -1.8105 & -0.6812 & -2.0716 & -0.9666 & -3.4306 \\ 
\bottomrule
\end{tabular}
\caption{Estimated parameters with average log-likelihood $\ell$ for the standard ACD model based on the generalized gamma distribution across the six exotic pairs.}
\label{tab:standard_exotic}
\end{table}

\begin{table}
\centering
\begin{tabular}{lcccccc}
\toprule
& USD/SGD & USD/KRW & USD/SEK & USD/CZK & USD/RON & USD/EGP \\
\midrule
$\omega$  & 1.0293 & 0.5951 & -0.9643 & 0.9209 & 0.2185 & -1.8846 \\ 
$\varphi$ & 0.9954 & 0.9910 & 0.9949 & 0.9976 & 0.9991 & 0.9701 \\ 
$\alpha$  & 0.0382 & 0.1358 & 0.0311 & 0.0440 & 0.0053 & 0.4452 \\ 
$\gamma$  & 0.7999 & 0.7017 & 1.2956 & 0.9944 & 0.7534 & 3.7676 \\ 
$\kappa$  & 1.0443 & 0.6475 & 0.8936 & 0.8730 & 1.3148 & 0.3452 \\ 
$\rho$    & 0.2111 & 0.8347 & 0.3476 & 0.2973 & 0.0875 & 0.0454 \\ 
$\sigma$  & 0.0185 & 0.2166 & 0.0998 & 0.0189 & 0.0320 & 0.0260 \\ \midrule
$\ell$    & -1.6431 & -1.3712 & -0.5032 & -1.6966 & -0.9243 & -3.4196 \\
\bottomrule
\end{tabular}
\caption{Estimated parameters with average log-likelihood $\ell$ for the granularity-adjusted ACD model based on the mixture distribution across the six exotic pairs.}
\label{tab:mixture_exotic}
\end{table}

\end{document}